\documentclass[11pt]{article}
\usepackage[utf8]{inputenc}
\usepackage{graphicx}
\usepackage{longtable}
\usepackage{hyperref}
\usepackage{amsmath}
\usepackage{amssymb}
\usepackage[margin=1in]{geometry}
\usepackage{natbib}
\usepackage{float}
\usepackage{booktabs}

\providecommand{\tightlist}{%
  \setlength{\itemsep}{0pt}\setlength{\parskip}{0pt}}

\title{Measuring Data Quality for Project Lighthouse\footnote{Any mistakes in this paper are the sole responsibility of Airbnb. In the interest of encouraging others to adopt this methodology, Airbnb and the authors formally disavow any intent to enforce their copyright in the content of this technical paper. Airbnb will not be liable for any indemnification of claims of intellectual property infringement made against users by third parties. We would like to thank Dr. Latanya Sweeney (Data Privacy Lab at Harvard) for early, confidential review.}}
\author{
Adam Bloomston\footnote{Corresponding Author; they may be reached at antidiscrimination-papers@airbnb.com.}, Elizabeth Burke, Megan Cacace, Anne Diaz, \\
Wren Dougherty, Matthew Gonzalez, Remington Gregg, Yeliz Güngör, Bryce Hayes, \\
Eeway Hsu, Oron Israeli, Heesoo Kim, Sara Kwasnick, Joanne Lacsina, \\
Demma Rosa Rodriguez, Adam Schiller, Whitney Schumacher, Jessica Simon, \\
Maggie Tang, Skyler Wharton, Marilyn Wilcken
}
\date{}

\begin{document}
\maketitle

\section{\texorpdfstring{{Introduction}}{Introduction}}\label{h.9obm8rdeslks}

{Airbnb continues to invest in making travel more open to everyone
\cite{Airbnb2024}. }{In 2020, we began using a tool called
Project Lighthouse, which was developed in partnership with leading
civil rights and privacy organizations, to help us uncover and address
potential disparities in how users of different perceived races may
experience Airbnb. }{Airbnb has used Project Lighthouse to evaluate
guests' booking success rate \cite{Airbnb2024}. In connection with that
work, we performed a simulation-based power analysis to ensure we could
accurately measure that experience gap under anonymization \cite{Basu2020}. As the variety of product experiences we may measure has
expanded, so too have our methods for assessing whether we can do so
under anonymization. This technical paper covers how we do so, by
measuring the preservation of data quality under anonymization.}

{}

{In this paper, we first situate the challenges for measuring data
quality under Project Lighthouse in the broader academic context. We
then discuss in detail the three core data quality metrics we use for
measurement---two of which extend prior academic work. Using those data
quality metrics as examples, we propose a framework, based on machine
learning classification, for empirically justifying the choice of data
quality metrics and their associated minimum thresholds. Finally we
outline how these methods enable us to rigorously meet the principle of
data minimization when analyzing potential experience gaps under Project
Lighthouse, which we term }{\textit{quantitative data minimization}}{.}

\section{\texorpdfstring{{Background}}{Background}}\label{h.h0s0wubhvt1c}

{Project Lighthouse was developed in close collaboration with experts
and partners to help us identify potential disparities in how users of
different perceived races experience the platform, while mitigating the
risk of sensitive attribute disclosure at scale. }{We met the goal of
preventing sensitive attribute disclosure through the application of two
distinct technical privacy models, k-anonymity and p-sensitive
k-anonymity, with one-way dataflows to enforce trust boundaries. We met
the goal of measuring potential }{experience}{~gaps by building a
bespoke prototype to run simulation-based power analyses that
demonstrated that, even under anonymization, we could measure
disparities in booking success rates with sufficient statistical power
\cite{Basu2020}.}

{}

{As Project Lighthouse matured, rather than investing in custom
simulation-based power analyses, we focused instead on measuring the
data quality (or, conversely, the loss of data quality) under
anonymization, i.e. in applying the technical privacy models. The
remainder of this technical paper focuses on the first technical privacy
model, k-anonymity, which has the largest impact on data quality \cite{Basu2020}.}

{}

{Much of the literature on data quality measurements for anonymization,
namely Privacy Preserving Data Mining (PPDM) and Privacy Preserving Data
Publishing (PPDP), presumes that the Publisher is }{\textit{Trusted}}{, so that
the Publisher has access to the identifiable sensitive data and may
compare it to the anonymized sensitive data to measure the impact of
anonymization \cite{Fung2010,Mendes2017,Carvalho2023}. That is not
the case for Project Lighthouse, so that we may measure data quality
under anonymization on identifiable, but not sensitive, data. In
contrast to PPDP, for Project Lighthouse there are an enumerable number
of statistical analyses an analyst may perform under Project
Lighthouse---so that the problem of measuring data quality is more
constrained than PPDP \cite{Basu2020}.}

{}

{The measures of data quality in the literature, which we call }{\textit{data
quality
metrics}}\footnote{\cite{Fung2010} calls these data metrics---we add the term quality for increased clarity. We are looking to measure the impact of a process, e.g. anonymization, on data quality, i.e. compare a dataset before and after a process---not assess the quality of a dataset in isolation; this maps to information loss, rather than information quality, in \cite{Fletcher2014}.}{~in
this paper, are usually proposed and used to compare two anonymization
algorithms, or various parameters for a single anonymization algorithm,
using the same dataset. The needs for analysis under Project Lighthouse,
and in Industry in general, are more complex---and will be addressed in
the next section.}

\section{\texorpdfstring{{D}{ata quality metrics for Project
Lighthouse}}{Data quality metrics for Project Lighthouse}}\label{h.ryxkj8rbbzvd}

{Our goal is to measure the impact of anonymization for an enumerable
number of potential analyses using only the identifiable non-sensitive
(e.g., without perceived race) and anonymized non-sensitive data. }{In
selecting and refining data quality metrics, we had four goals:}

{}

\begin{enumerate}
\tightlist
\item
  {They ought to be constrained to a specific interval and
  interpretable. In our case we chose {[}0, 1{]} where 0 reflects
  minimal data quality and 1 reflects maximal data quality.}
\item
  {They ought to be comparable across columns, i.e. a value is produced
  for each column and it is sensible to compare values between two
  columns to say that one column has ``superior'' data quality under
  anonymization to another. This helps guide the analyst when data
  quality is insufficient.}
\item
  {They ought to be comparable across complex datasets. We analyze
  datasets of varying row cardinality, column cardinality, column types,
  etc.}
\item
  {Finally, and building on the three goals above, they ought to be
  useful for risk decisions (most importantly, quantitative data
  minimization, which is covered in a later section) made by analysts
  who may not have expertise in anonymization nor private data
  analysis.}
\end{enumerate}

{}

{Below we outline the primary three data quality metrics we use that we
have found to satisfy goals 1 and 2 above; then we discuss how we derive
a single value for each data quality metric for a dataset (as opposed to
each column in that dataset) to satisfy goals 3 and 4 above. Finally we
account for limitations in these data quality metrics by supplementing
them with a fourth, secondary data quality metric on the entire
dataset.}

\subsection{\texorpdfstring{{Terminology and example
dataset}}{Terminology and example dataset}}\label{h.2flf36kplqvd}

{For discussion below, assume we have an original dataset $O$ whose
$i$-th column values are represented by the vector $O_i$. And it
has been anonymized as $A$, the non-suppressed records, and $S$, the
suppressed records; with $A_i$ and $S_i$ the $i$-th column values.
Let $e$ be an equivalence class such that the union of all $e$ is
$A$ and $e_i$ be the anonymized (i.e. generalized,
micro-aggregated) attribute values for the $i$-th quasi-identifier for
all records in $e$. For convenience when a formula operates on
$O_i$, $A_i$ assume that only the associated non-suppressed
records in $O$ are considered so that the vectors have equal length.
Each column $i$ may either be numerical or categorical in type; a
binary (True/False) column is considered a numerical, and the 0/1
values may be micro-aggregated under anonymization to be in {[}0, 1{]},
and thus interpreted probabilistically for analysis \cite{Basu2020}.}

{}

{For the first two data quality metrics discussed below, we will also
use the example dataset from \cite{Basu2020}, replicated in the two
tables below; the third table is a joined version of the first two, for
easy comparison between $O$ and $A$ for later use.}

\begin{figure}[H]
\centering
\includegraphics[width=0.45\textwidth,keepaspectratio]{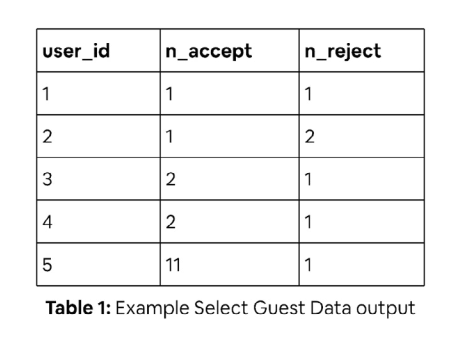}
\hfill
\includegraphics[width=0.45\textwidth,keepaspectratio]{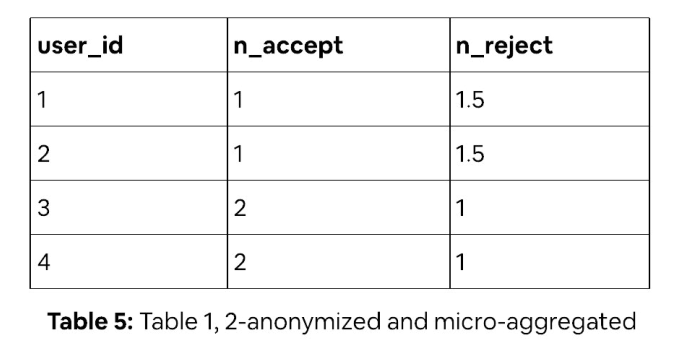}
\caption{Example dataset from Basu 2020: Original data (left) and anonymized data (right) showing user acceptance and rejection counts before and after k-anonymity is applied.}
\label{fig:example-dataset}
\end{figure}

{}

\begin{longtable}[]{@{}lllll@{}}
\toprule\noalign{}
\endhead
\bottomrule\noalign{}
\endlastfoot
{user\_id} & {n\_accept\_orig} & {n\_reject\_orig} & {n\_accept\_anon} &
{n\_reject\_anon} \\
{1} & {1} & {1} & {1} & {1.5} \\
{2} & {1} & {2} & {1} & {1.5} \\
{3} & {2} & {1} & {2} & {1} \\
{4} & {2} & {1} & {2} & {1} \\
{5} & {11} & {1} & {\textless suppressed\textgreater{}} &
{\textless suppressed\textgreater{}} \\
\end{longtable}

\subsection{\texorpdfstring{{Pearson's correlation
coefficient}}{Pearson's correlation coefficient}}\label{h.592jpocit5j3}

{Our first data quality metric is }{Pearson's correlation coefficient}{,
a commonly used data quality metric \cite{Kim2011}. For each numerical
column $i$, it is the Pearson's correlation coefficient between
$O_i$ and $A_i$, ignoring missing
values}\footnote{In a subsequent paper we will outline the anonymization algorithm we use, which handles missing values.}{,
and floored at 0:}

{}

{RHO(O, A, i) = max{[}0, cov($O_i$, $A_i$) / (std($O_i$) *
std($A_i$)){]}}

{}

{A value of 1.0 indicates that there is a perfect positive linear
relationship between the original and anonymized data for a numerical
column; and a value 0.0 indicates that either no linear relationship
($\rho$ = 0) or a negative linear relationship ($\rho$ \textless{} 0).
This data quality metric may be thought of as measuring the
preservation of linear relationships under anonymization. Note that we
may have e.g. $A_i$ \textasciitilde{} 2 * $O_i$ + 3 and yet
RHO(O, A, i) = 1.0, i.e. a perfect linear relationship does not
imply a $\beta$ (slope) of 1.0 nor an $\alpha$ (intercept) of 0.0 for line fit;
in practice, however, our anonymization algorithm (to be discussed in a
subsequent technical paper) as designed does not mutate the data in such
a way. In future work, however, we may modify this data quality metric
to account for such a possibility.}

{}

{For the example dataset, we have }{RHO(O, A, n\_accept) = 1.00}{~and
}{RHO(O, A, n\_reject)}{~= 0.58:}

\begin{figure}[H]
\centering
\includegraphics[width=0.45\textwidth,keepaspectratio]{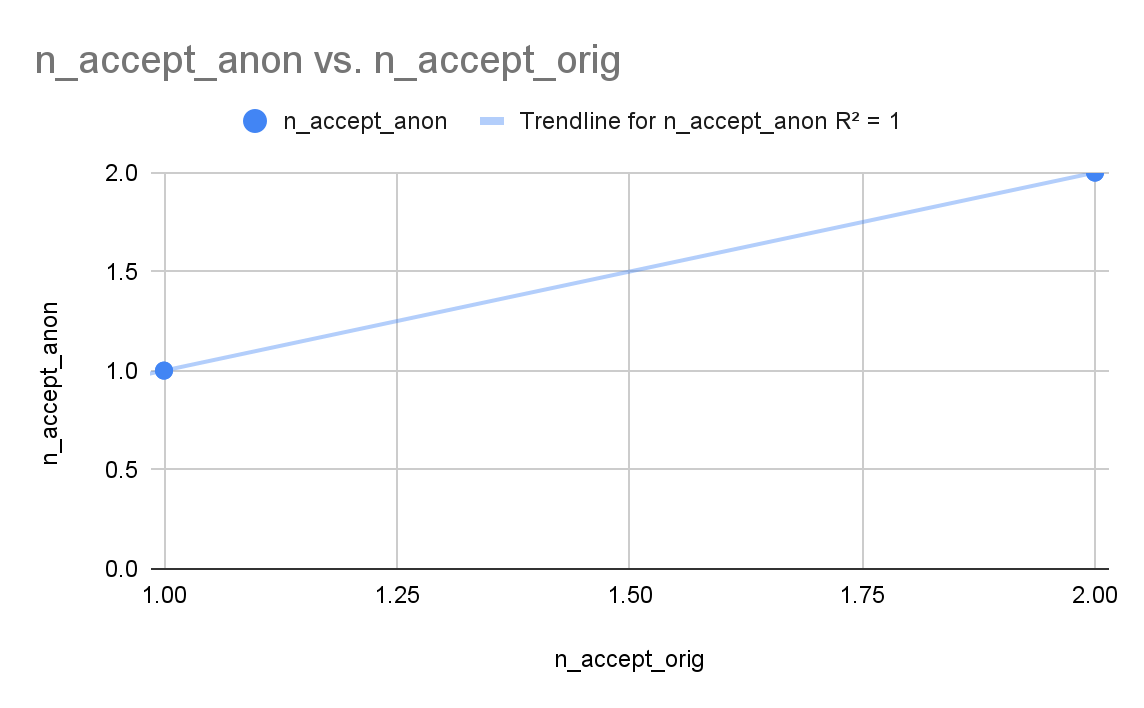}
\hfill
\includegraphics[width=0.45\textwidth,keepaspectratio]{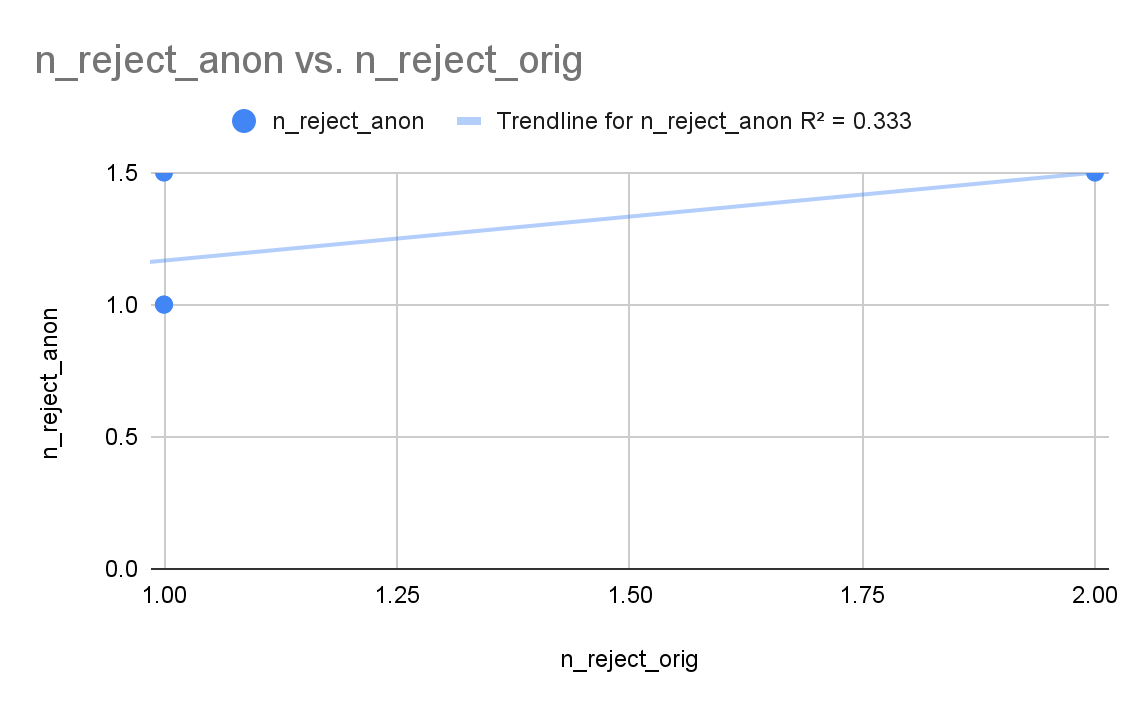}
\caption{Pearson's correlation coefficient for the example dataset: n\_accept shows perfect correlation (RHO = 1.00, left) while n\_reject shows moderate correlation (RHO = 0.58, right) between original and anonymized values.}
\label{fig:pearson-correlation-plots}
\end{figure}

\subsection{\texorpdfstring{{Revised Information Loss
Metric}}{Revised Information Loss Metric}}\label{h.pcrf4n8976gj}

{Our second data quality metric is the }{Revised Information Loss
Metric}{~(RILM), a revision and extension of the Information Loss Metric
(ILM) from \cite{Byun2006}}\footnote{RILM may also be considered as a revision and extension of the Classification Metric (CM) from \cite{Iyengar2002} where the penalty function has been modified, and suppression ignored; however, we encountered ILM before CM---so maintain the terminology based on ILM.}{.
This data quality metric may be thought of as measuring }{\textit{the
preservation of geometric size under anonymization}}{. It has been
revised to be a column-level score, rather than a dataset level score;
and extended to cover categorical columns. ILM is built on IL, which is
defined on each equivalence class e in A. If we rewrite
}{\textit{\textbar G}}{j}{\textit{\textbar{}}}{, }{\textit{\textbar D}}{j}{\textit{\textbar{}}}{~from
Definition 7 of \cite{Byun2006} as
$\textit{perim}(e_i)$, $\textit{perim}(O_i)$:

{}

$\textit{perim}(O_i) = \max(O_i) - \min(O_i)$

$\textit{perim}(e_i) = \max(e_i) - \min(e_i)$

{}

{Then IL and ILM may be written, using the terminology adopted for this
section, as:}

{}

$\textit{IL}(e) = |e| \times \Sigma \{ \textit{perim}(e_i)/\textit{perim}(O_i) \text{ if } \textit{perim}(O_i) > 0 \text{ else } 0 \}$ for all numerical quasi-identifiers $i$

{ILM(O, A) = $\Sigma$ IL(e) / \textbar A\textbar{} for all equivalence classes
e in A}

{}

\subsubsection{RILM for numericals}

{We define RILM for numericals by modifying IL,
ILM be column-level (goal 2), and subtracting from 1 (goal 1):}

{}

$\textit{RIL}(e, i) = \textit{perim}(e_i)/\textit{perim}(O_i) \text{ if } \textit{perim}(O_i) > 0 \text{ else } 0$

{\textit{RILM(O, A, i)= 1 - \{($\Sigma$ \textbar e\textbar{} * RIL(e, i)~)/
\textbar A\textbar\} for all equivalence classes e in A}}

{}

{If the perimeter (max - min) of all equivalence classes for
quasi-identifier $i$ is unchanged, then we have $\textit{perim}(e_i) = 0$,
}{\textit{RILM(e, i) = 0}}{, and thus }{\textit{RILM(O, A, i) = 1 - 0 = 1}}{. If, however,
there is a single equivalence class with all values, and there are
at-least two distinct values, then }{\textit{RIL(e, i) = 1/1 = 1}}{~and }{\textit{RILM(O,
A, i) = 1 - 1 = 0}}{.}

{}

{For the example dataset, we have }{\textit{RILM(}}{n\_accept}{\textit{)}}{~\textit{= 1.00}}{~and
}{\textit{RILM(}}{n\_reject}{\textit{)}}{~\textit{= 0.50}}{:}

{}

\begin{longtable}[]{@{}ll@{}}
\toprule\noalign{}
\endhead
\bottomrule\noalign{}
\endlastfoot
{perim(n\_accept)} & {10.00} \\
{perim(n\_reject)} & {1.00} \\
\end{longtable}

{}

\begin{longtable}[]{@{}lllll@{}}
\toprule\noalign{}
\endhead
\bottomrule\noalign{}
\endlastfoot
{equivalence class} & {perim(n\_accept)} & {RIL(n\_accept)} &
{perim(n\_reject)} & {RIL(n\_reject)} \\
{user\_id in (1, 2)} & {0.00} & {0.00} & {1.00} & {1.00} \\
{user\_id in (3, 4)} & {0.00} & {0.00} & {0.00} & {0.00} \\
\end{longtable}

{}

\subsubsection{RILM for categoricals}

{For each categorical quasi-identifier, we
define a generalization hierarchy, or }{\textit{g-tree}}{, for the purposes of
generalization \cite{Samarati1998}. For example, suppose we have three
original values for a categorical: ``foo'', ``bar'', and ``test''; we
may define a g-tree as follows:}

{}

\begin{figure}[H]
\centering
\includegraphics[width=0.6\textwidth,keepaspectratio]{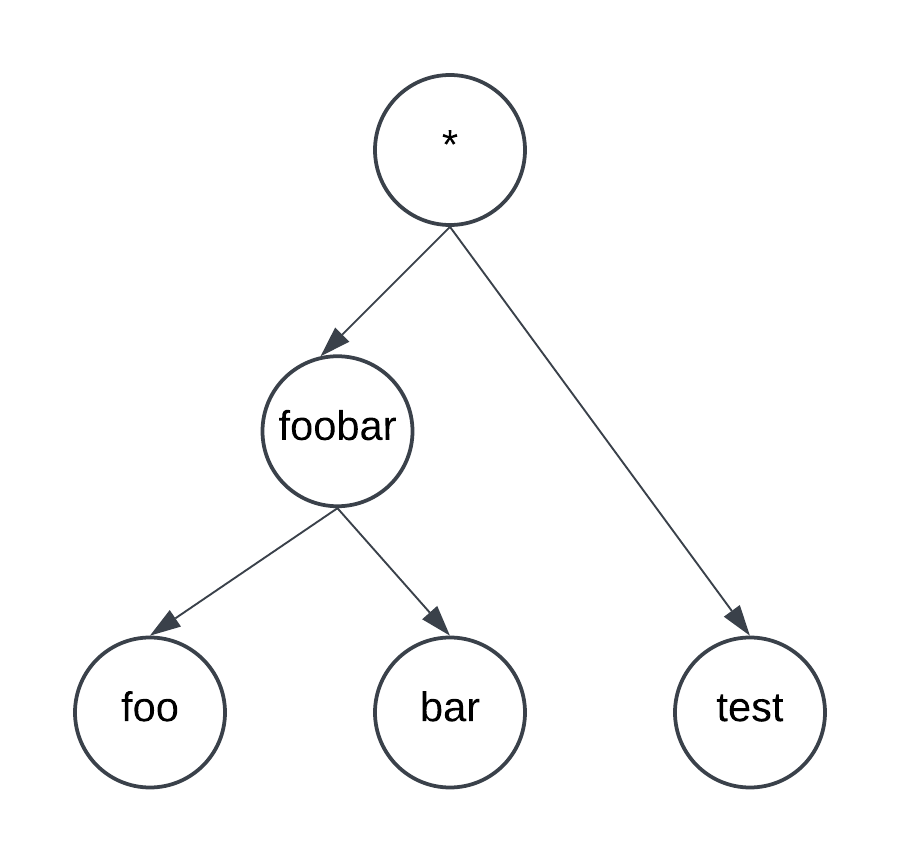}
\caption{Example generalization tree (g-tree) structure for categorical values showing how ``foo'' and ``bar'' can be generalized to ``foobar'', and all three values generalize to the root node ``*''.}
\label{fig:gtree-structure}
\end{figure}

{To extend RILM for categorical quasi-identifiers, we mandate that each
g-tree has ``geometric sizes'' associated with each node in the g-tree,
with non-decreasing sizes as we move up towards the root and all leaf
nodes having a value of 0.0. Using our example above, we may have:}

{}

\begin{figure}[H]
\centering
\includegraphics[width=0.6\textwidth,keepaspectratio]{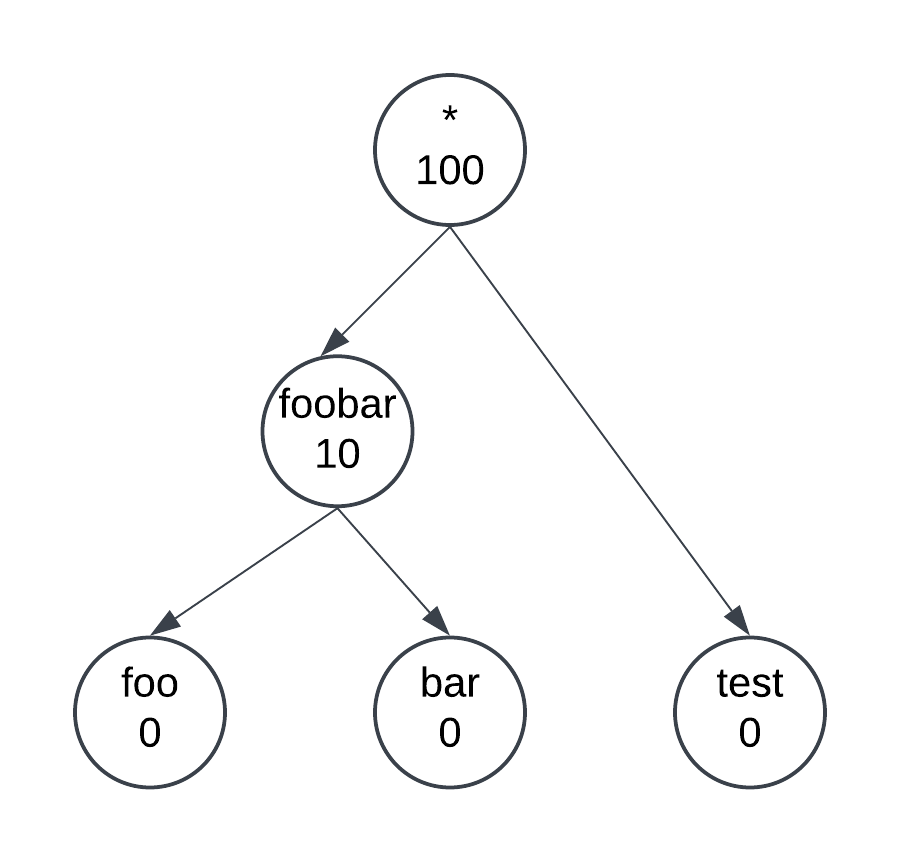}
\caption{Generalization tree with geometric sizes: Leaf nodes (foo, bar, test) have size 0.0, intermediate node (foobar) has size 10.0, and root node (*) has size 100.0, used to calculate RILM for categorical quasi-identifiers.}
\label{fig:gtree-geometric-sizes}
\end{figure}

{}

{And finally we define }{\textit{perim}}{~for categorical quasi-identifiers,
which is the only change in computing RILM for categoricals vs.
numericals, as:}

{}

$\textit{perim}(O_i) = \text{geometric size of root, *}$

$\textit{perim}(e_i) = \text{geometric size of generalized node in g-tree for } e_i$

{}

{For example, suppose we have an equivalence class whose original values
were }{\textit{foo}}{, }{\textit{bar}}{~and were generalized to }{\textit{foobar}}{. Then we have:}

{}

$\textit{perim}(O_i) = 100$

$\textit{perim}(e_i) = 10$

{}

{So that }{\textit{RIL(e, i) = 10 / 100 = 0.10}}{. Suppose instead that the
original values were }{\textit{foo}}{, }{\textit{bar}}{, }{\textit{test}}{~so that the generalized
value must be *; then:}

{}

$\textit{perim}(O_i) = 100$

$\textit{perim}(e_i) = 100$

{}

{So that }{\textit{RIL(e, i) = 10 / 100 = 1.0}}{.}

{}

{For most categorical quasi-identifiers we do not have a complicated
g-tree}\footnote{In subsequent technical papers we will provide more complicated use cases requiring g-trees.}{,
instead we automatically generate a default, }{\textit{flat g-tree}}{~where the
depth is 2, all unique original values are leaf nodes with geometric
size 0, and the root node }{\textit{*}}{~has some geometric size \textgreater{}
0. Then RILM simply becomes a measure of local cell suppression---higher
local or cell suppression yields a lower RILM, as expected \cite{Fung2010}. Because ``not all generalization steps are created equal,'' we
create custom g-trees, with geometric sizes determined by a
domain-expert, for commonly used categorical quasi-identifiers \cite{Kifer2006}.}

{}

{In a subsequent technical paper, we will show how we use an extension
of RILM (both numerical and categorical) as a search metric, using the
terminology from \cite{Fung2010}, for our anonymization algorithm.}

\subsection{\texorpdfstring{{Normalized Mutual Information v1, Sampled
and
Scaled}}{Normalized Mutual Information v1, Sampled and Scaled}}\label{h.20jmmrca7rhi}

{Our third data quality metric is }{Normalized Mutual Information v1,
Sampled and Scaled (NMIv1)}{. }{Mutual information}{~(MI) measures the
mutual dependence between two variables. This data quality metric may be
thought of as measuring }{the preservation of minimal entropy under
anonymization}{. In contrast to Pearson's coefficient, MI provides a
more general measure that doesn't assume a linear relationship between
the two variables. In this subsection, we first introduce the
}{\textit{normalization}}{~divisor (for goal 1); we then introduce a random model
for discussion and use it to introduce }{\textit{scaling}}{, and finally
}{\textit{sampling}}{.}

{}

{Let MI($O_i$, $A_i$) be the mutual information between the
original and anonymized values for numerical quasi-identifier
$i$---note that it ranges in {[}0, $\infty$). We \textit{normalize} to {[}0, 1{]}
in one of two ways, both leveraging that MI(X, Y) \textless= H(X),
H(Y) where H(X) is the entropy of X \cite{McDaid2013,Vinh2009}:}

{}

{NMIv1(O, A, i) = MI($O_i$, $A_i$) / H($O_i$)}

{NMIv2(O, A, i) = MI($O_i$, $A_i$) / H($O_i$)}

{}

{A high value of NMIv1 means that most of the information in the
original data is present in the anonymized data, whereas a high value of
NMIv2 means that most of the information in the anonymized data is
present in the original data. In other words, NMIv2 predominantly
penalizes the injection of entropy by the anonymization process, whereas
NMIv1 predominantly penalizes the suppression of entropy by the
anonymization process. ~See example 2 in the next section for a
discussion of the choice of NMIv1 over NMIv2---the remainder of this
subsection focus on NMIv1.}

{}

{For the purposes of discussion in this sub-section we introduce the
following random model:}

{}

{Let }{X}{, }{Y, Z}{~be three independent random variables drawn from
}{U{[}0,1{]}}{; let's construct a model }{t = X + bXY + cZ}{~with }{b}{,
}{c}{~parameters in {[}0, 1{]} We may interpret the dependent variable t
and independent random variables X, Y, Z as follows:}

{}

\begin{itemize}
\tightlist
\item
  {X}{~represents the original input data for a specific numerical
  quasi-identifier.}
\item
  {bXY}{~represents some mutation that occurs to }{X}{~under
  anonymization that is a function of }{X}{---this might be interpreted
  as mutation ``local'' to the quasi-identifier in question.
  }{b}{~represents the magnitude of this impact---a larger }{b}{~might
  represent a larger }{k}{~in the case of k-anonymity,}{~or increased
  variance in values of the specific quasi-identifier---both of which
  would increase the impact of local perturbation.}
\item
  {cZ}{~represents some mutation that occurs to }{X}{~under
  anonymization (e.g. micro-aggregation following \cite{Basu2020}) that
  is independent of the value of }{X}{, due e.g. to ``global''
  perturbation because of another independent quasi-identifier (i.e. not
  }{X}{). }{c}{~represents the magnitude of this impact---a larger
  }{c}{~might represent more independent quasi-identifiers, a larger
  }{k}{~for k-anonymization, or increased variance of the independent
  quasi-identifiers.}
\item
  {Finally }{t}{~represents the output, anonymized data, for that same
  quasi-identifier.}
\end{itemize}

{}

{For discussion below we select }{b = 0.40}{, }{c = 0.05}{, and scale
}{X}{~to }{X \textasciitilde{} 1000.0 * U{[}0, 1{]}}{. And to modify the
entropy in }{X}{~we may round it---which will be represented as e.g.
}{round(1000.0 * U{[}0, 1{]}, -2)}{~to round to the nearest value of
100.}

{}

{When analyzing real data for potential experience gaps, we may have
vastly different entropy in the original data. In practice, we have
found that the information needed for analyzing those potential
experience gaps does not scale linearly to the input entropy,}{~i.e. as
the entropy increases less of the total entropy is necessary for
successful analysis and thus must be preserved under
anonymization.}{~But NMIv1 as currently defined is just the ratio of
}{\textit{nats}}{, or units of entropy, preserved under anonymization; to account
for this, we }{\textit{scale}}{~NMIv1 to reduce the cost of losing nats as the
original input entropy increases---the intuition here is that
information lost is likely lost across all original nats equally, and
some nats are more important than others and, the less original nats,
the more likely we have lost from important nats. We scale NMIv1 through
an exponential penalty, whereby each successive nat in the input data
receives an increasingly smaller penalty. Let $n$ be the value of
NMIv1(O, A, i) as defined above, i.e. before scaling, $e$ be the input
entropy H($O_i$) (also in nats); then the scaled version is:}

{}

\begin{equation}
\text{NMIv1}(O, A, i) = 1 - \frac{1}{e} \int_0^e \frac{1}{2^x} (1 - n) \, dx
\end{equation}

{The approximate version may help elucidate how the penalty (}{n}{) is
being applied to each original nat, were we to assume an integral number
of nats, with the penalty decreasing exponentially with each successive
nat:}

{}

\begin{equation}
\text{NMIv1}(O, A, i) \approx 1 - \frac{1}{e} \sum_{x=0}^{e} \frac{1}{2^x} (1 - n)
\end{equation}

{}

{So that e.g. the first nat has a penalty of }{1 * (1 - n)}{, the second
has a penalty of }{(½) * (1 - n)}{, etc.}

{}

{We can explore this further using the random model, with four levels of
rounding and associated plots:}

{}

\begin{figure}[H]
\centering
\includegraphics[width=0.45\textwidth,keepaspectratio]{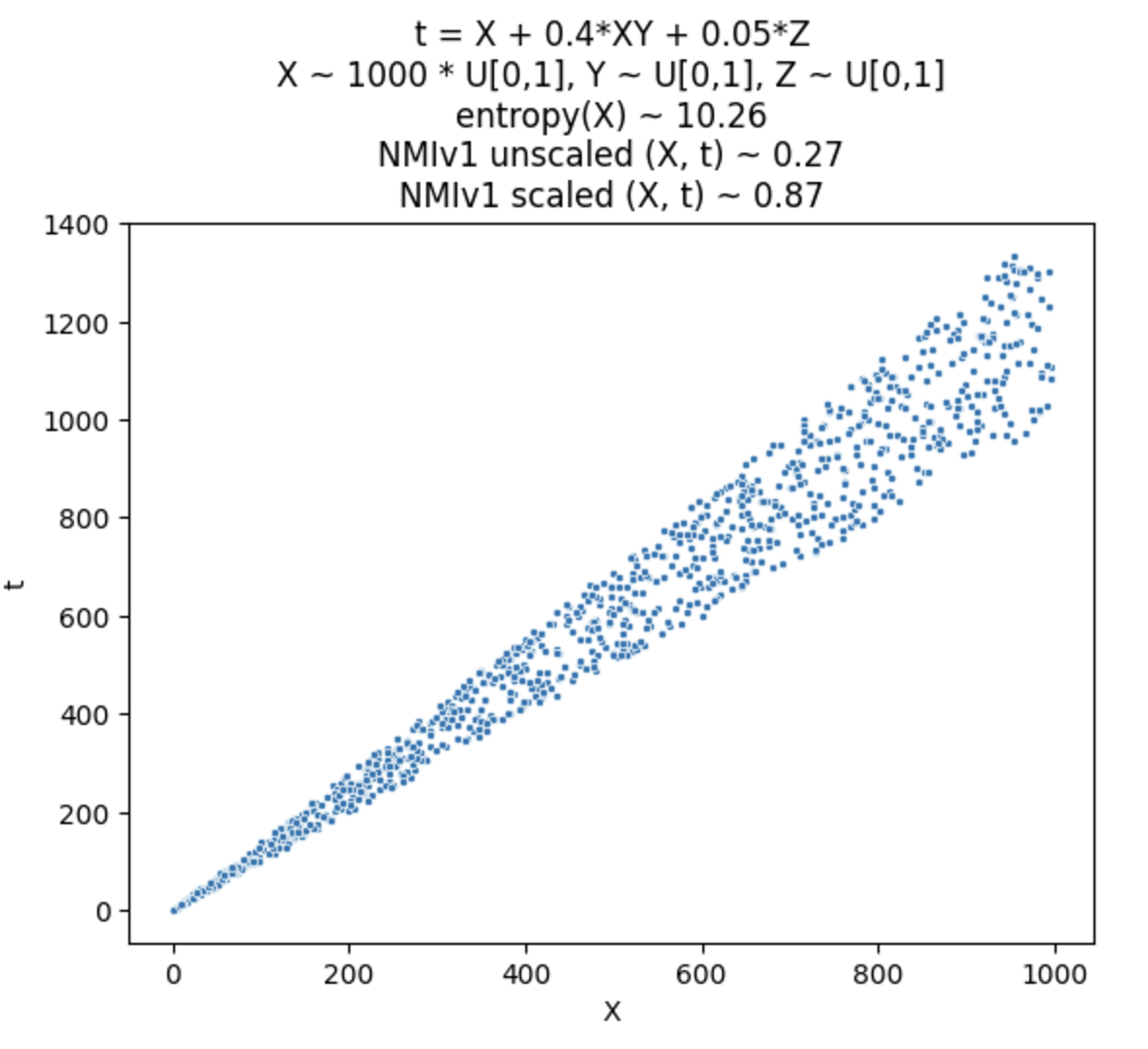}
\hfill
\includegraphics[width=0.45\textwidth,keepaspectratio]{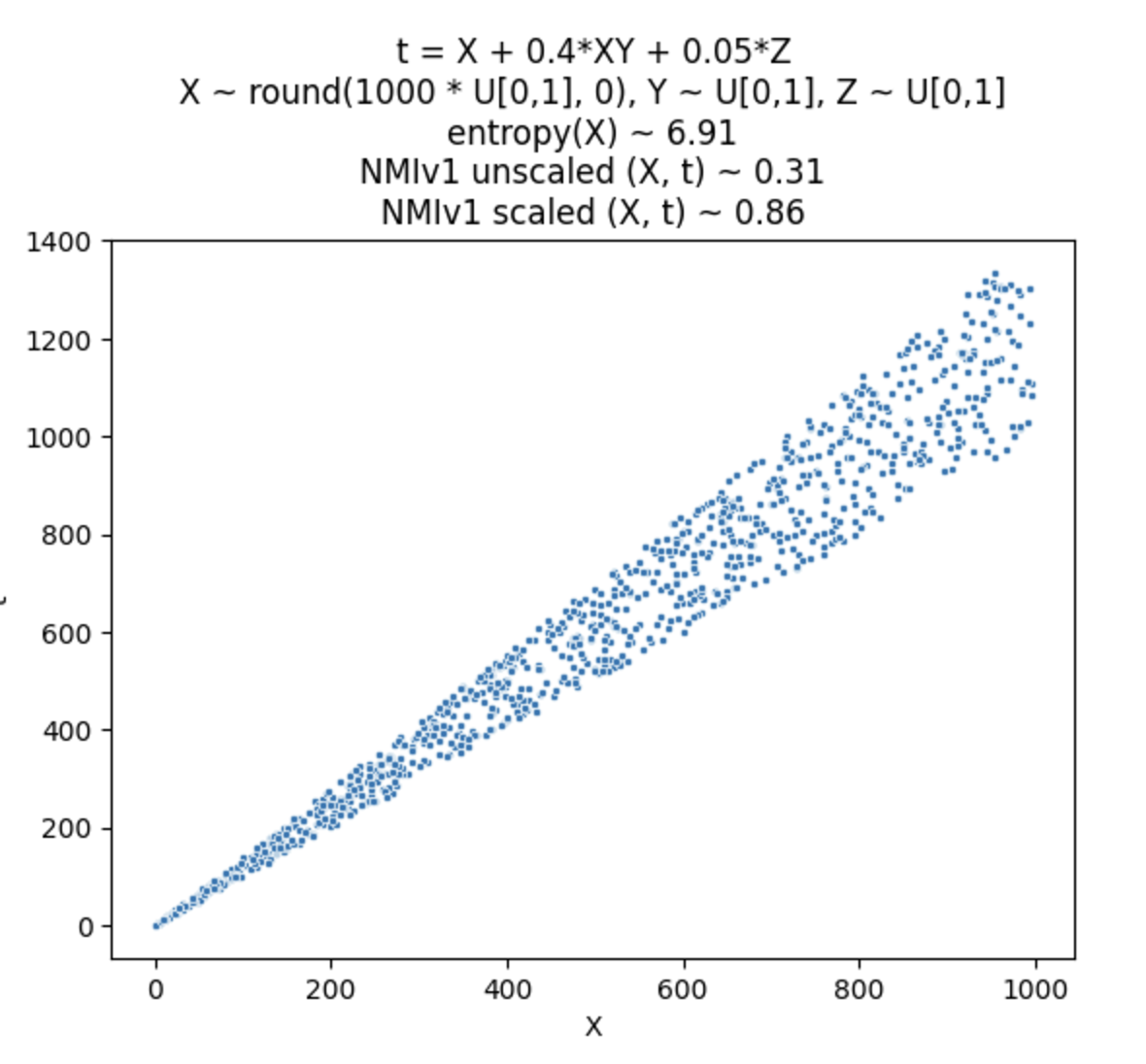}
\\[1em]
\includegraphics[width=0.45\textwidth,keepaspectratio]{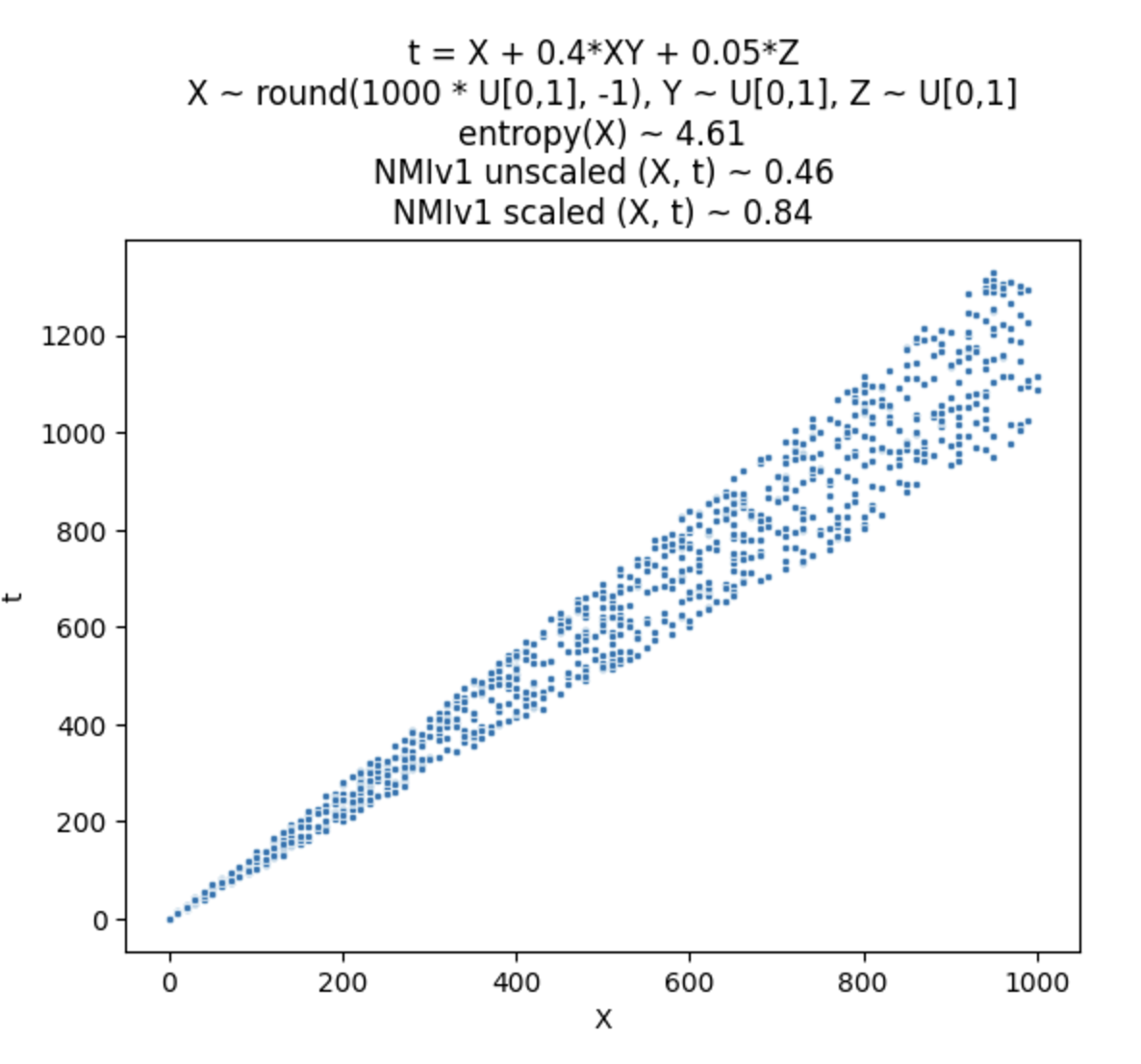}
\hfill
\includegraphics[width=0.45\textwidth,keepaspectratio]{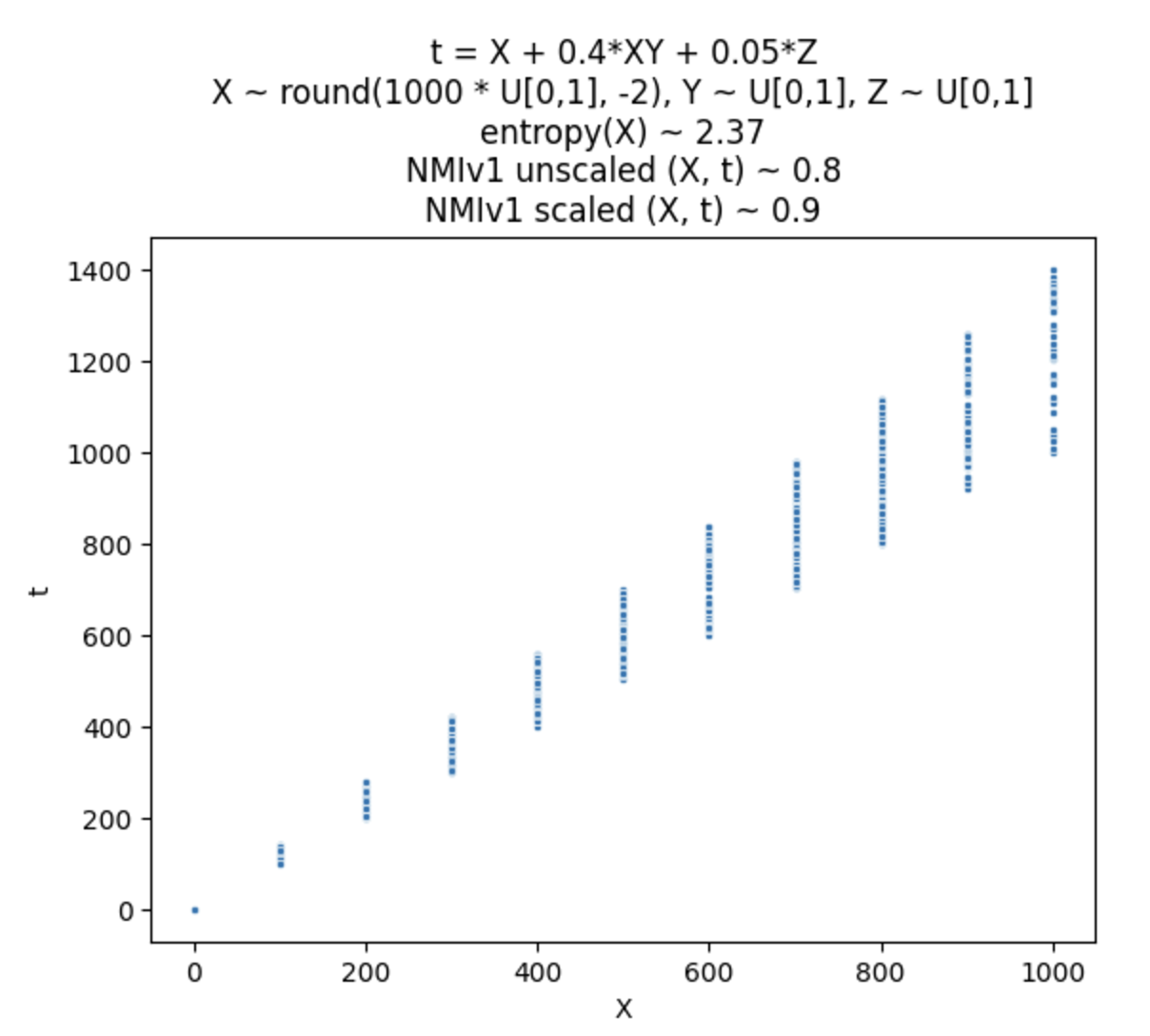}
\caption{NMIv1 scaling demonstration with random model at four entropy levels: no rounding (H=10.26 nats, top left), round to nearest integer (H=6.91 nats, top right), round to nearest 10 (H=4.61 nats, bottom left), and round to nearest 100 (H=2.37 nats, bottom right). Scaled NMIv1 produces comparable scores (0.84-0.90) across all plots despite vastly different unscaled scores (0.27-0.80).}
\label{fig:nmiv1-scaling-demonstration}
\end{figure}

{No rounding is used in the first plot, and the entropy of }{X}{~(for
discussion, consider }{X}{~as the original data) is \textasciitilde10.26
nats; for the second plot }{X}{~is rounded to the nearest integer, i.e.
}{round(.., 0)}{, which reduces the original data entropy to
\textasciitilde6.91 nats; for the third }{X}{~is rounded to the nearest
value of 10, for an entropy of \textasciitilde4.61 nats; finally, for
the fourth plot }{X}{~is rounded to the nearest value of 100, for an
entropy of \textasciitilde2.37 nats.}

{}

{If we examine these four plots, intuitively we see the general trend
(as }{X}{~increases }{t}{~increases linearly, but the noise also
increases) preserved so that we wish to consider a comparable amount of
}{\textit{minimal}}{~information to be preserved across at-least the first three
plots. However, the data quality metric NMIv1 unscaled has vastly
different scores for them---ranging from 0.27 (the first plot) to 0.8
(the fourth plot). If we scale NMIv1, i.e. exponentially reduce the
penalty as the original data entropy (}{X}{) increases, then we have
}{\textit{comparable}}{~scores across all four plots---ranging from 0.84 to
0.90.}

{}

{For computing mutual information and entropy, we rely on the
implementations in scikit-learn; in practice we see the estimate
asymptotically approaching what appears to be the actual entropy /
mutual information as the sample size used for computing entropy /
mutual information increases \cite{Pedregosa2011,Kraskov2004}. This
may be due to the MI estimator requiring a large sample size for
accurate estimation---which we may explore in future work, but for now
accept as given \cite{Gao2015}. The plot below demonstrates this for the
random model above.}

{}

\begin{figure}[H]
\centering
\includegraphics[width=0.8\textwidth,keepaspectratio]{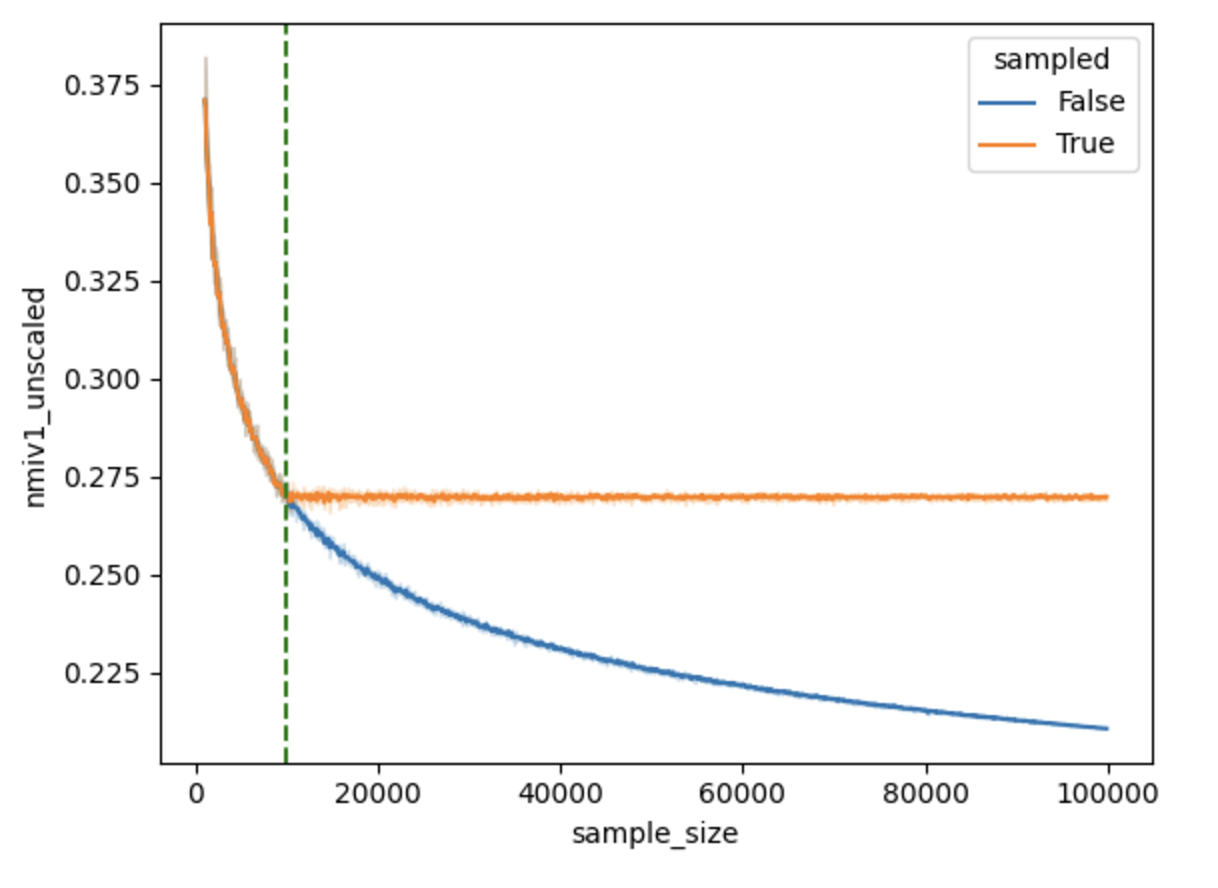}
\caption{NMIv1 sampling behavior: Blue curve (sampled=False) shows asymptotic convergence to ~0.21 as sample size increases. Orange curve (sampled=True) follows the same curve until reaching 10k records, then stabilizes at ~0.27, ensuring comparable NMIv1 values across datasets of different sizes.}
\label{fig:nmiv1-sampling}
\end{figure}

{The blue curve (for }{sampled = False}{) shows the NMIv1 value
asymptotically approaching \textasciitilde0.21 as the sample size used
for computing mutual information (and thus entropy) increases. Following
goal 3 above, we wish to compare NMIv1 across datasets of varying
sizes---we would not, for instance like to see NMIv1 = 0.225 for a
dataset of \textasciitilde60k records and, for a sample of 10k records
from that same dataset, see NMIv1 = 0.275, or a \textasciitilde22\%
increase. Because the exact value of NMIv1 is less important than its
ability to satisfy the goals above, and because in practice we are
examining a relatively large number records, when we have at-least 10k
records we pseudo-randomly \textit{sample} 10,000 values for $O_i$,
$A_i$ a number of times, compute NMIv1 for each, and take the average
across those samples. The orange curve (for }{sampled = True}{) shows
the sampled NMIv1 value following the same asymptotic curve as the blue
curve (for }{sampled = False}{) until it reaches the minimum size for
sampling to be enabled---and then it hovers around \textasciitilde0.27.
So that 10k and 100k of records from the same dataset should have
comparable values of NMIv1.}

{}

{In future work we may extend this data quality metric to account for
sophisticated models of information in the anonymized data \cite{Goldberger2009}.}

\subsection{\texorpdfstring{{Percent of non-suppressed
records}}{Percent of non-suppressed records}}\label{h.p2k74yypld3f}

{The above primary data quality metrics do not account for suppressed
records; in future work we may extend them to be appropriately penalized
by suppression. In the meantime we supplement them with a fourth data
quality metric, }{the }{percent of non-suppressed records}{:}

{}

{\textit{PCTNS = \textbar A\textbar{} / \textbar O\textbar{} = 1.0 -
\textbar S\textbar{} / \textbar O\textbar{}}}

{}

{A value of 1.0 indicates no suppression, and a value of 0.0 indicates
that all records have been suppressed. In the example dataset we have
}{\textit{PCTNS = 4 / 5 = 0.80}}{.}

\subsection{\texorpdfstring{{Minimum data quality metric
values}}{Minimum data quality metric values}}\label{h.q5zjntp4mhla}

{To satisfy our third goal above, we would like to have a set of data
quality metric values for comparison across datasets with different
columns. Because we assume all columns are equally important for
analysis, we take the minimum of the column-level values of each of the
primary three data quality metrics. }{In the example dataset above, we
have }{\textit{RILM(n\_accept) = 1.00}}{~and }{\textit{RILM(n\_reject) = 0.50}}{---so that
the dataset level RILM score is }{\textit{RILM = MIN(RILM(n\_accept),
RILM(n\_reject)) = MIN(1.00, 0.50) = 0.50}}{.}{~}{For the fourth data
quality metric, we already have a single value.}

{}

{To satisfy our fourth goal, we define a minimum threshold for each data
quality metric. If the minimum data quality metric value, for each data
quality metric, meets or exceeds its associated threshold, we say that
the A }{\textit{meets minimum data quality}}{.}

{}

{Through conservative trial and error, we have arrived at the data
quality metric thresholds that work best for our specific needs---these
are described in a later section. As part of that process, we developed
a framework for comparing data quality metrics and selecting appropriate
thresholds for them---this process and its results are described in the
next section.}

\section{\texorpdfstring{{Empirical}{~justification for data quality
metrics}}{Empirical~justification for data quality metrics}}\label{h.diir7x16lq8a}

{The literature proposes many data quality metrics, whose justification
may be theoretically grounded in the underlying tasks for which the data
are being constructed, e.g. \cite{Kifer2006}, or may be minimally
justified and instead used for a specific comparison of e.g. two
anonymization algorithms. Critically, when data quality metrics are
introduced in the literature, little to no guidance is given regarding
appropriate thresholds to determine if/when anonymization is ``good
enough.'' In this section, we offer an empirical methodology for
comparing data quality metrics and determining appropriate thresholds
for them; it may be tuned for particular applications through the
selection of appropriate datasets and statistical tests. Finally, we end
this section with examples that show how this methodology may be used
for making decisions about data quality metrics.}

\subsection{\texorpdfstring{{Methodology}}{Methodology}}\label{h.9jakl8cw1pm1}

{The goal of a data quality metric is to tell the analyst whether the
data are ``good enough'' after anonymization for a particular use. In
the case of Project Lighthouse, this determination is made without
reference to sensitive attributes---but in this section }{we use public
datasets}{~where those sensitive attributes may be used for the purposes
of determining appropriate data quality metrics and thresholds.}

{}

\begin{figure}[H]
\centering
\includegraphics[width=0.8\textwidth,keepaspectratio]{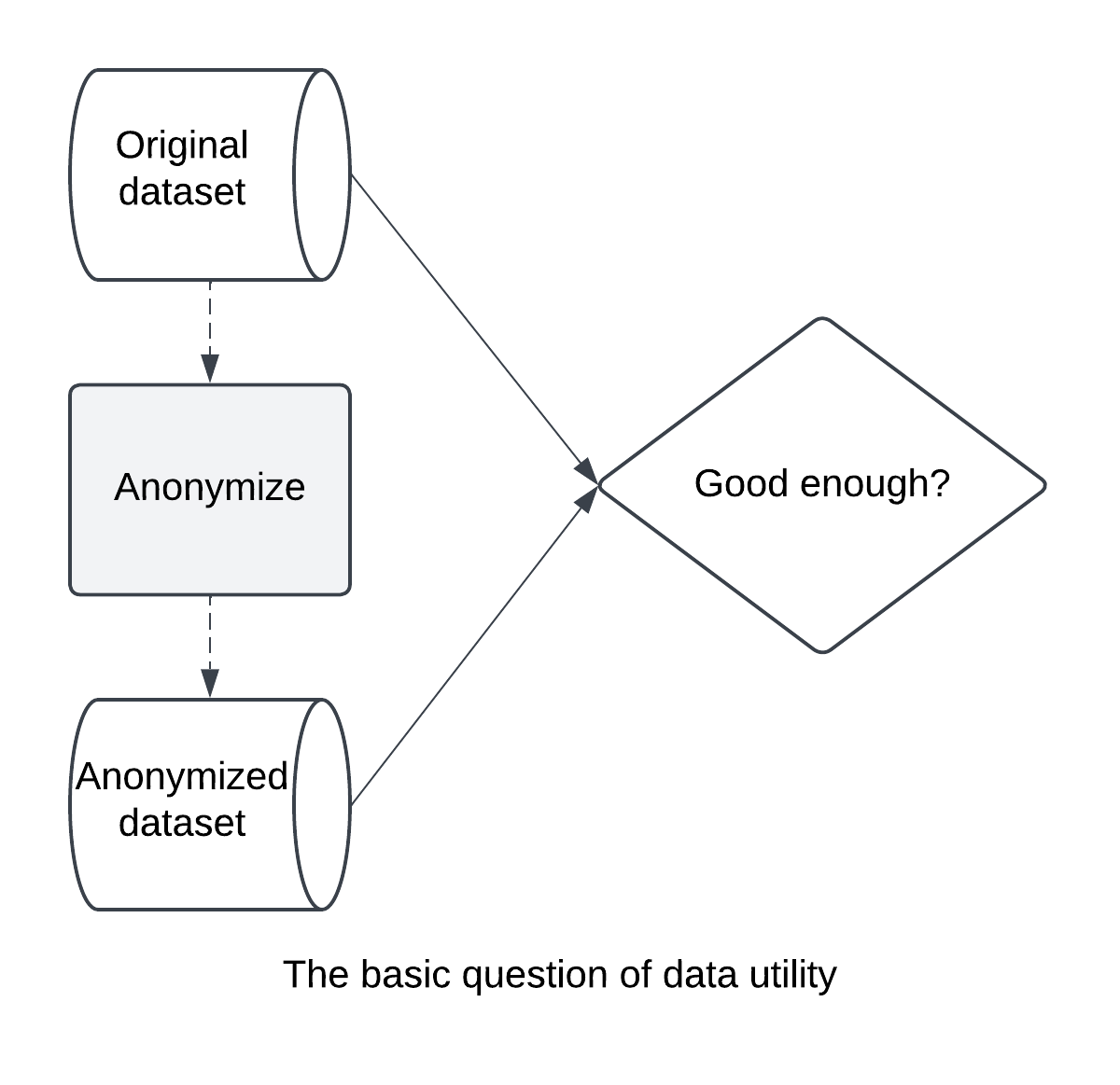}
\caption{Empirical justification methodology overview: Data quality metrics are evaluated as classifiers that predict whether anonymized datasets are ``good enough'' for analysis, using public datasets where sensitive attributes enable validation.}
\label{fig:methodology-overview}
\end{figure}

{}

{We may think of the data quality metrics as ``classifiers'' that
determine if anonymized datasets are ``}{good enough'' for
analysis}{---where ``good enough'' is defined below. We will thus
reframe the problem of justifying data quality metrics as a machine
learning classification problem, and utilize the common methodologies
for such problems. In that framing, the goal is to find a model and
threshold that together may predict whether a given anonymized dataset
is ``good enough''.}

{}

\begin{figure}[H]
\centering
\includegraphics[width=0.8\textwidth,keepaspectratio]{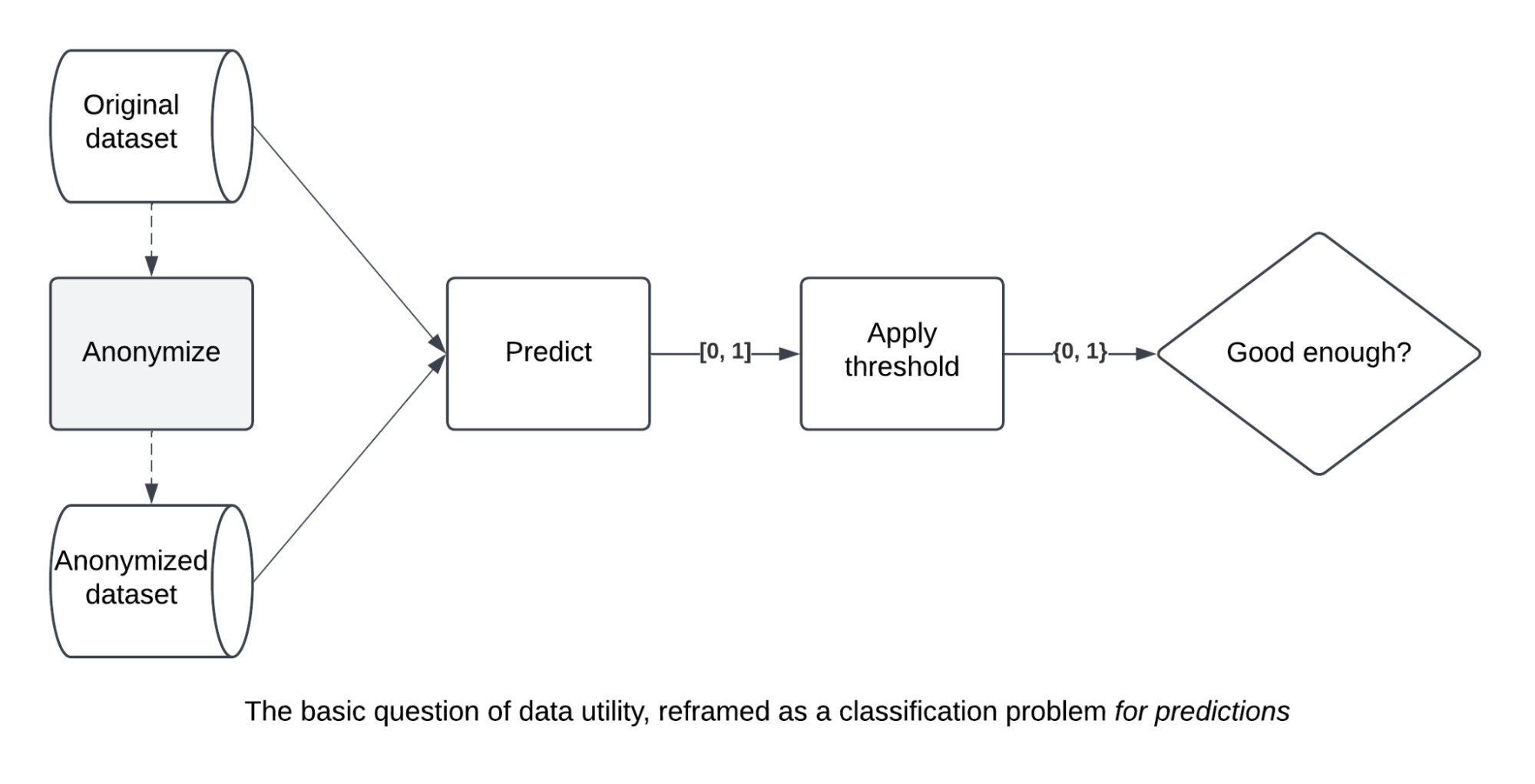}
\caption{Classification framework for data quality metrics: Reframing the validation problem as a machine learning classification task where data quality metrics serve as models and thresholds determine whether anonymized datasets are predicted to be ``good enough'' for analysis.}
\label{fig:classification-framework}
\end{figure}

{In this framing, we require labels that answer if an anonymized dataset
is ``good enough''; we will produce these labels by comparing the output
of relevant statistical tests between the original and anonymized
datasets---if we consider the statistical tests on the original dataset
as ground truth, then the result of the statistical test on the
anonymized dataset may be compared to that ground truth and its error
classified as one of the following:}

{}

\begin{longtable}[]{@{}lll@{}}
\toprule\noalign{}
\endhead
\bottomrule\noalign{}
\endlastfoot
{} & {Original dataset statistical test} & {Anonymized dataset
statistical test} \\
{Type I error} & {Statistical significance} & {No statistical
significance} \\
{Type II error} & {No statistical significance} & {Statistical
significance} \\
{Test statistic sign mismatch} & {Statistical significance} &
{Statistical significance} \\
\end{longtable}

{}

{If they match in statistical significance and test statistic sign then
the anonymized dataset is considered ``good enough'' (}{label = 1}{),
otherwise it is not (}{label = 0}{).}

{}

\begin{figure}[H]
\centering
\includegraphics[width=0.8\textwidth,keepaspectratio]{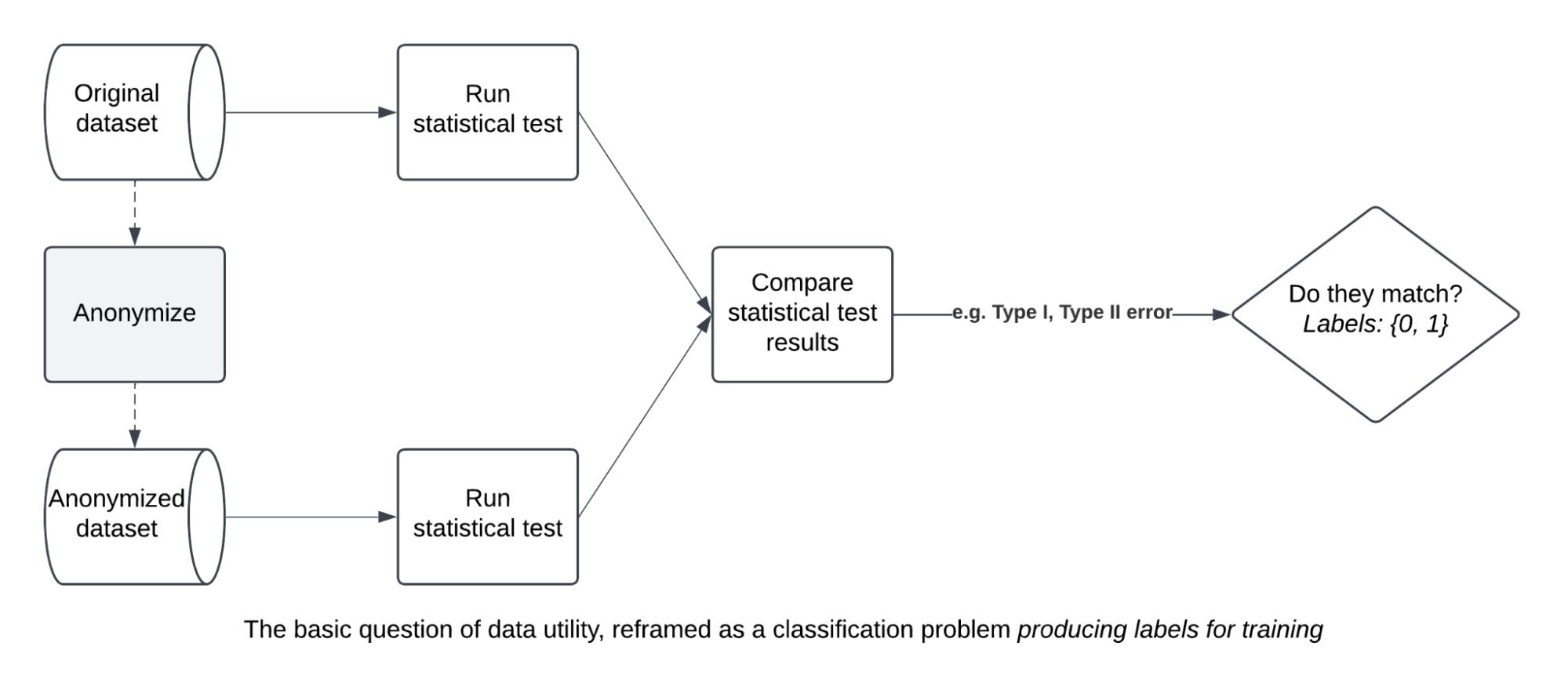}
\caption{Label generation for classification: Anonymized datasets are labeled as ``good enough'' (label=1) when statistical tests match the original in both significance and sign, otherwise labeled as insufficient (label=0).}
\label{fig:label-generation}
\end{figure}

{}

{Following others' work, we utilize the Adults public Census dataset
\cite{Becker1996,Ghinita2007,LeFevre2006,Machanavajjhala2006}. In
future work we may utilize additional datasets and/or mutate those
datasets for our particular needs. We will consider, in line with
\cite{Basu2020}, all data columns as quasi-identifiers except
}{race}{---}{which will be considered the sensitive attribute. For
statistical tests we will look for potential experience gaps with
respect to }{race}{---for categorical columns a maximum likelihood
G-test, and for numerical columns a t-test. In future work, we may
consider additional statistical tests, including those that involve
multiple columns.}

{}

{In order for us to properly assess the efficacy of our ``models,'' we
wish to test them on a large number of labels, i.e. have a large number
of }{(Original dataset, Anonymized dataset)}{~pairs from which we can
produce labels as described above. To produce those datasets, we first
utilize a GAN to produce a large number of synthetic, original datasets
whose characteristics are in line with the original Adults public Census
dataset \cite{Xu2019}. Then, for each synthetic dataset we may anonymize
it in a variety of ways (e.g. different values of k for k-anonymity),
producing multiple }{(Original synthetic dataset, Anonymized
dataset)}{~pairs. And each time we anonymize we will also compute the
data quality metrics described in the prior section---those are the
predicted scores when considering this as a classification problem.}

{}

\begin{figure}[H]
\centering
\includegraphics[width=0.8\textwidth,keepaspectratio]{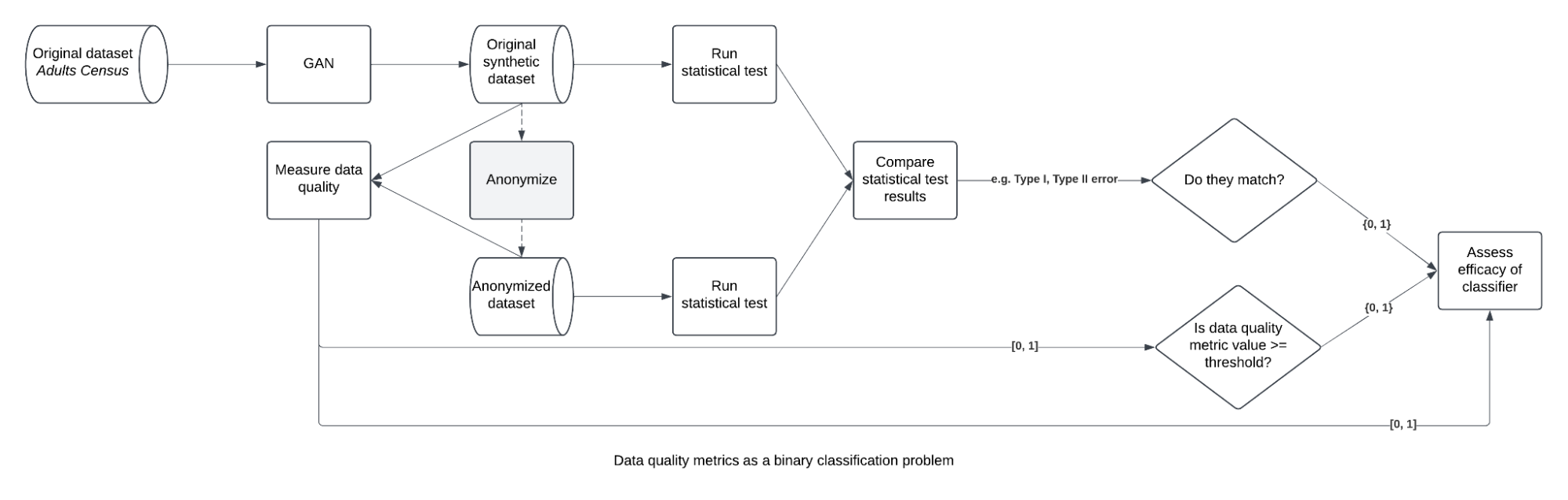}
\caption{Synthetic dataset generation and anonymization pipeline: GANs generate multiple synthetic datasets from the original Adults Census data, each anonymized with varying parameters (e.g., different k values) to produce numerous (Original, Anonymized) pairs with computed data quality metrics.}
\label{fig:gan-simulation-pipeline}
\end{figure}

{We now have the following for each quasi-identifier in }{(Original
synthetic dataset, Anonymized dataset)}{:}

{}

\begin{enumerate}
\tightlist
\item
  {The data quality metric scores associated with that
  quasi-identifier.}
\item
  {Labels: 0 or 1, for each demographic group.}
\end{enumerate}

{}

{Framed as a classification problem, we now have predicted scores and
labels for a number of models---each data quality metric (Pearson's,
RILM numerical, RILM categorical, NMIv1 Sampled Scaled) may be
considered a model. So that we may first assess the efficacy of those
models, and then may select appropriate thresholds for each. And our
measurements of efficacy, following standard approaches for
classification problems, may consider the accuracy of original patterns
}{and}{~the impact of erroneous artificial patterns \cite{Fletcher2014}.}

{}

{The original dataset had approximately 50,000 records, four numerical
columns (considered as quasi-identifiers), and eight categorical columns
(seven considered quasi-identifiers, the eighth }{race}{~considered the
sensitive attribute). After training a GAN, we produce synthetic
datasets of varying numbers of records less than or equal the number of
records in the original dataset to minimize erroneous statistical
significance as compared to the original, i.e. non-synthetic, dataset
\cite{Lin2013}. For each synthetic dataset we may select one or more of
the quasi-identifiers (columns besides race) for consideration. For each
synthetic dataset and quasi-identifier combination, we anonymize with
different values of k. For a trained GAN, then, we have approximately
20,000 distinct applications of anonymization. Because the GAN training
process can produce varying results, we train the GAN twenty times
}{independently and simulate results as described above for each trained
GAN, for a sum}{~total of approximately 400}{,000}{~distinct
applications of anonymization. }{Each application of anonymization is a
simulation run that produces labels for each (quasi-identifier, race
value) that is appropriate.}

\subsection{\texorpdfstring{{Results}}{Results}}\label{h.nduuz7ri1y0l}

{With this simulated data we may now assess the efficacy of our data
quality metrics (``models'') and examine possible thresholds we may
select. Two standard approaches to assessing the efficacy of a
classification model are to examine its ROC and Precision-Recall (PR)
curves---below we show those curves for each data quality metric. The
curves shown are computed on the aggregated results from all twenty
trained GANs. And for ROC we also plot the ROCs of the trained GAN that
produced the worst and best AUC; }{\textit{similarly}}{~for Average Precision and
PR curves.}

{}

{Each plot also indicates a suggested threshold: this is the smallest
threshold in {[}0.0, 1.0) such that the total percent of errors (total
Type I, Type II, mismatch in test statistic sign, divided by total
number of statistical tests) is \textless= 5\%.}

{}

{Each plot also shows the thresholds we use for Project Lighthouse
analyses, as described in \cite{Basu2020}, as of the publication of this
technical
paper}\footnote{Note that these thresholds have been chosen based upon extensive internal analysis and experience using anonymization for Project Lighthouse; only in the case of NMIv1 Sampled, Scaled did the empirical justification cause us to revise our minimum threshold (we revised it downward, from 0.90 to 0.80).}{,
summarized here:}

{}

\begin{itemize}
\tightlist
\item
  {Pearson's 0.90.}
\item
  {RILM for numericals: not used, but included in this paper for
  completeness.}
\item
  {RILM for categoricals 0.90; }{the empirical justification for
  \textless= 0.98 is lacking---which we believe is due to improvements
  needed in the simulation-based analysis.}{~In future work we will
  extend the statistical tests used to support custom g-trees, which
  should help elucidate the minimum threshold more clearly. Also note
  that there are a few special-case quasi-identifiers where we use a
  lower threshold (0.60)---these will be discussed in a subsequent
  technical paper.}
\item
  {Normalized Mutual Information v1, Sampled and Scaled: 0.80.}
\item
  {Percent of non-suppressed records: 0.99; this too lacks empirical
  justification in isolation; in future work we will extend these data
  quality metrics to account for suppression and remove this threshold.}
\end{itemize}

\begin{figure}[H]
\centering
\includegraphics[width=0.8\textwidth,keepaspectratio]{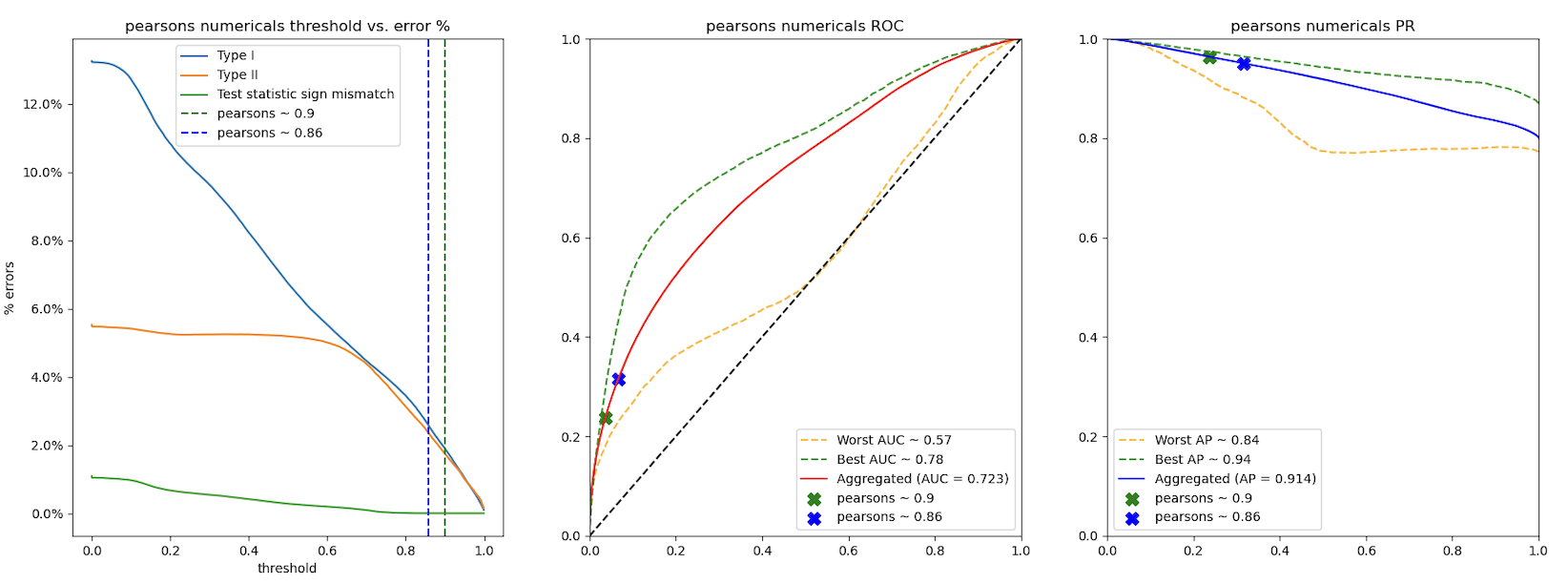}
\caption{Pearson's correlation coefficient: ROC and Precision-Recall curves showing classifier performance (AUC, AP) with suggested threshold and Project Lighthouse threshold (0.90) marked. Curves show best, worst, and aggregated performance across 20 trained GANs.}
\label{fig:results-pearson}
\end{figure}

\begin{figure}[H]
\centering
\includegraphics[width=0.8\textwidth,keepaspectratio]{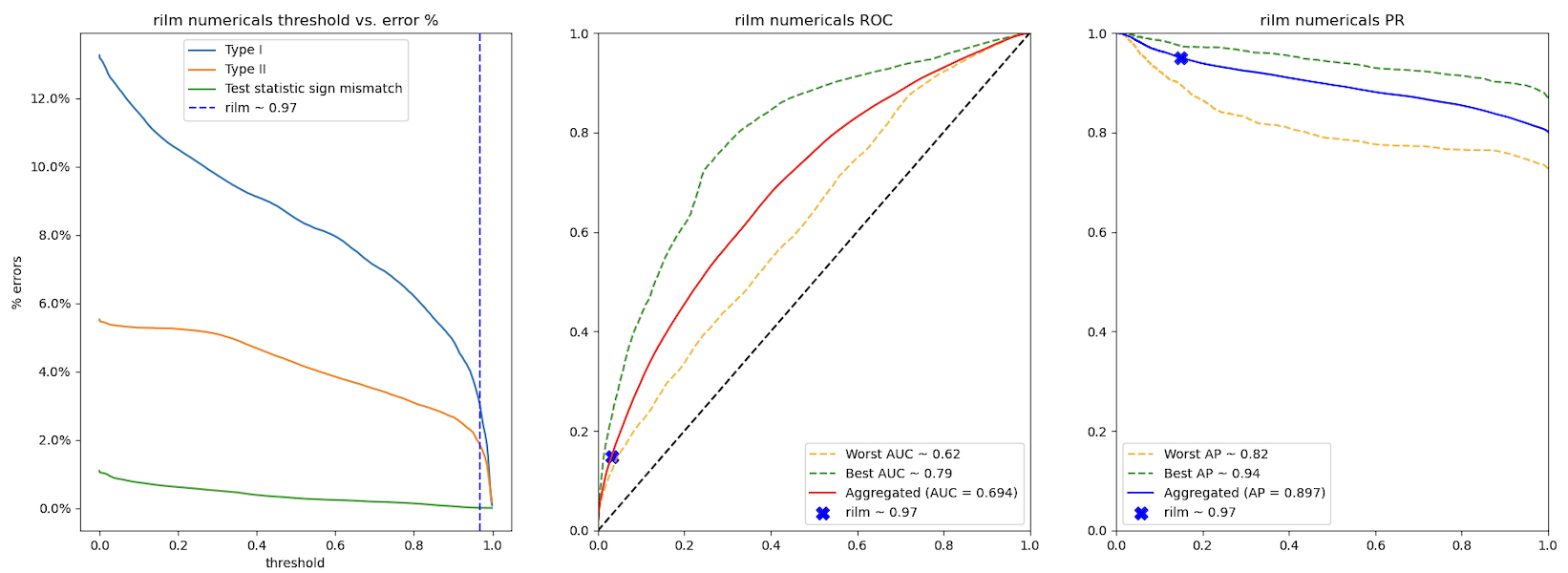}
\caption{RILM for numerical quasi-identifiers: ROC and Precision-Recall curves with performance metrics. While not currently used as a threshold in Project Lighthouse, included for completeness and comparison with ILM.}
\label{fig:results-rilm-numerical}
\end{figure}

\begin{figure}[H]
\centering
\includegraphics[width=0.8\textwidth,keepaspectratio]{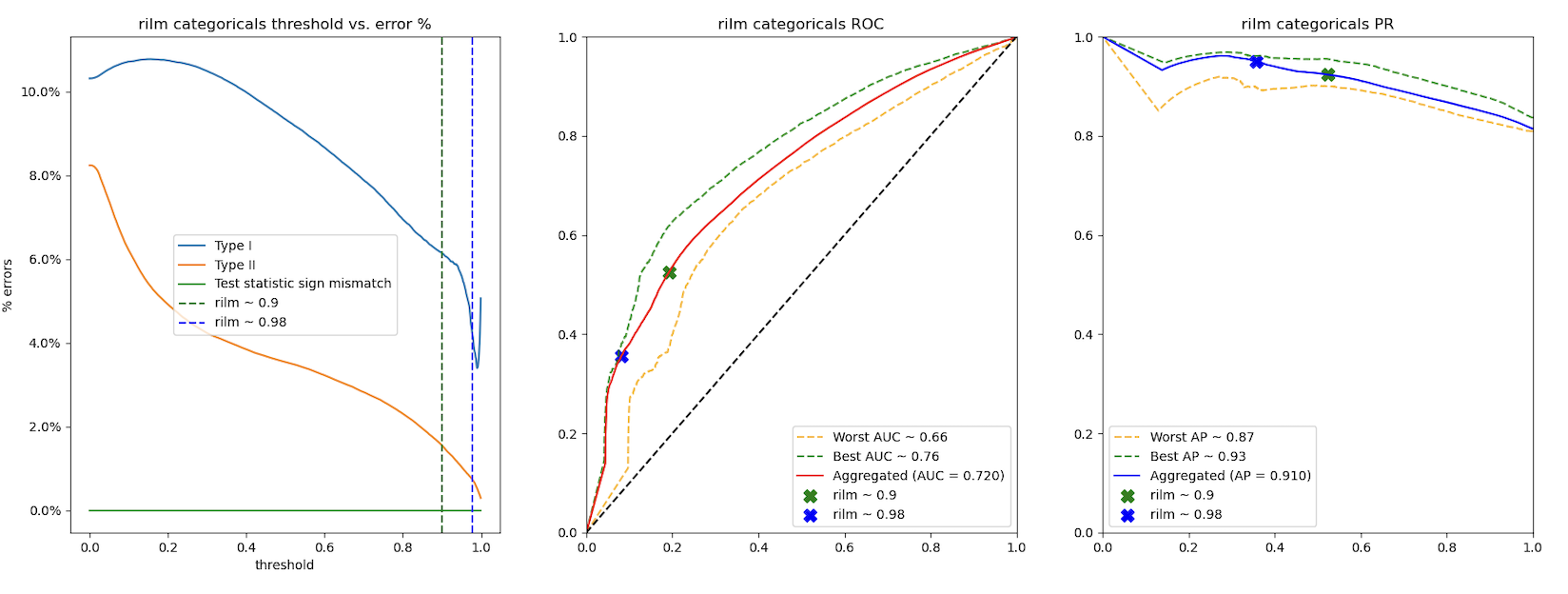}
\caption{RILM for categorical quasi-identifiers: ROC and Precision-Recall curves showing performance with Project Lighthouse threshold of 0.90. Note that empirical justification for thresholds $\leq$0.98 requires further investigation with enhanced statistical tests for custom g-trees.}
\label{fig:results-rilm-categorical}
\end{figure}

\begin{figure}[H]
\centering
\includegraphics[width=0.8\textwidth,keepaspectratio]{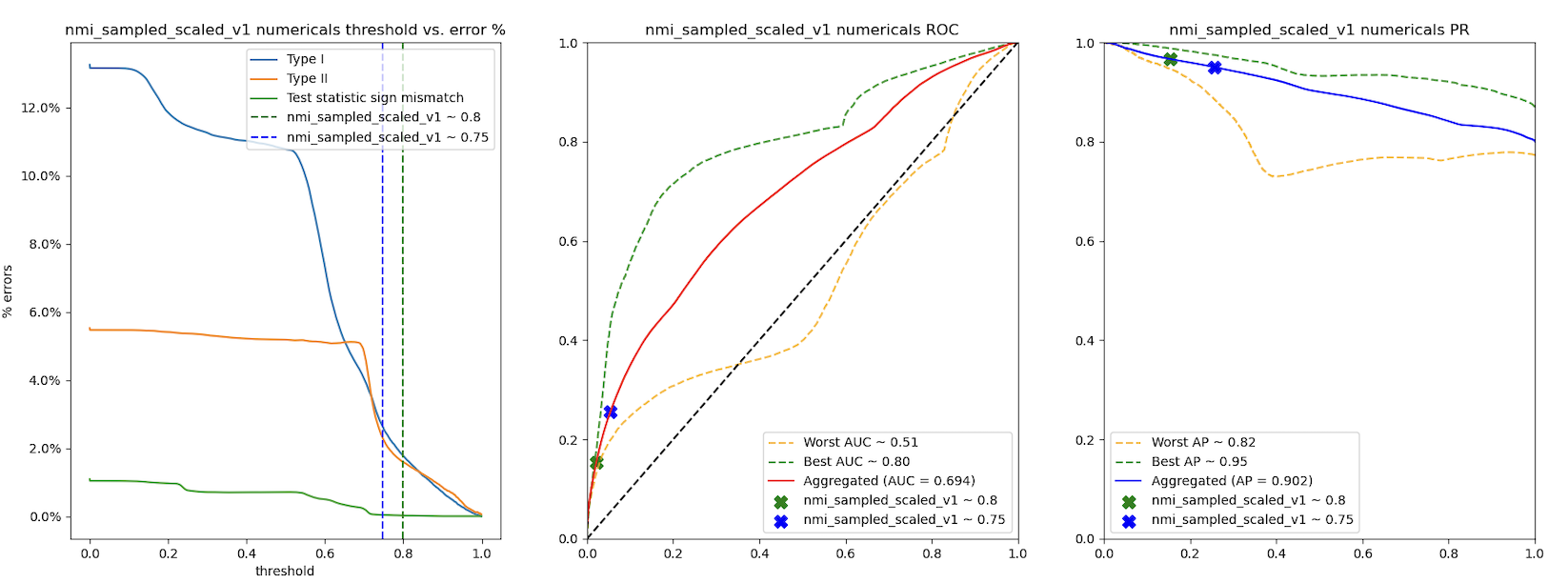}
\caption{Normalized Mutual Information v1, Sampled and Scaled: ROC and Precision-Recall curves with Project Lighthouse threshold of 0.80 (revised downward from initial 0.90 based on empirical justification).}
\label{fig:results-nmiv1}
\end{figure}

\begin{figure}[H]
\centering
\includegraphics[width=0.8\textwidth,keepaspectratio]{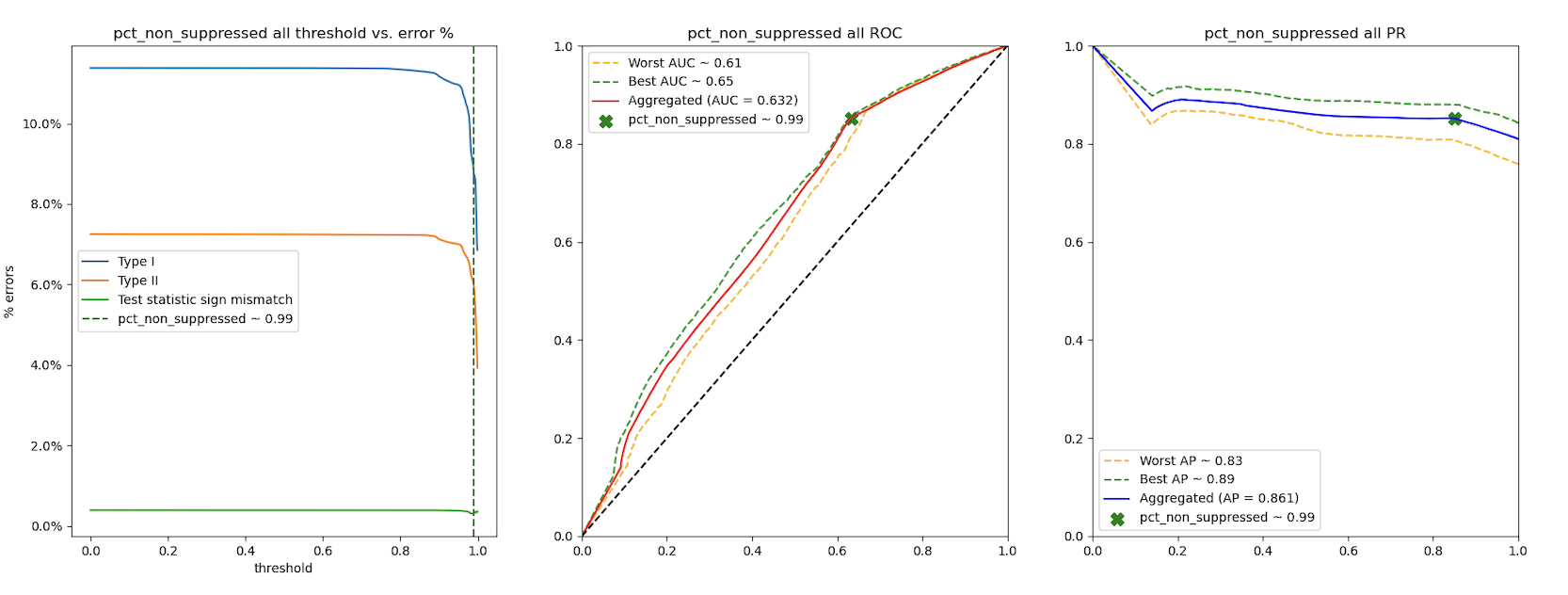}
\caption{Percent of non-suppressed records: ROC and Precision-Recall curves with Project Lighthouse threshold of 0.99. This metric currently lacks strong empirical justification in isolation; future work will integrate suppression penalties into primary data quality metrics.}
\label{fig:results-pctns}
\end{figure}

{}

{Below we show two examples where this classification framing helps to
investigate potential data quality metrics---these work under the
assumption that an appropriate original dataset (or datasets) and
statistical tests have been selected.}

{}

{Example 1.}{~In proposing RILM numerical as a revision on ILM one may
ask: is RILM an appropriate replacement for ILM? A common comparison
between two models is to compare their ROC Area Under the Curve (AUC)
and Average Precision (AP). The RILM numerical plot above may be
compared to the ILM plot below---we use }{\textit{1 - ILM}}{~to allow for easy
comparison}{.}

{}

\begin{figure}[H]
\centering
\includegraphics[width=0.8\textwidth,keepaspectratio]{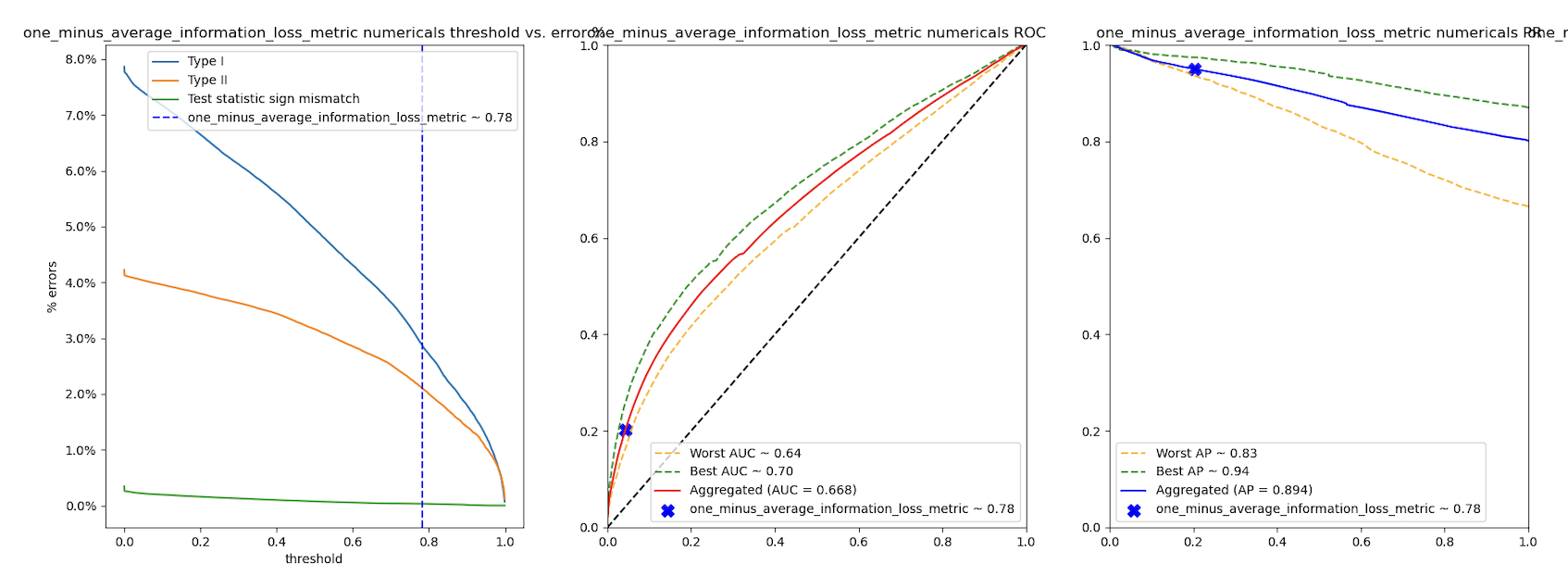}
\caption{Information Loss Metric (1-ILM) comparison: ROC and Precision-Recall curves for ILM showing comparable AUC and AP to RILM numerical, validating RILM as a reasonable column-level revision of the dataset-level ILM metric.}
\label{fig:example1-ilm}
\end{figure}

{In this case, the two models have comparable efficacy because their AUC
and AP are quite similar. This indicates that RILM may be a reasonable
substitute for ILM---though an in-depth comparison of results applying
all data quality metric thresholds may be warranted, and is outside the
scope of this technical paper.}

{}

{Example 2.}{~In defining NMI one may ask, which is a more appropriate
divisor to normalize the mutual information between the original and
anonymized values: the original values' entropy (NMIv1) or the
anonymized values' entropy (NMIv2)? Below we provide the v2 plots, for
comparison with the NMIv1 plots given above.}

{}

\begin{figure}[H]
\centering
\includegraphics[width=0.8\textwidth,keepaspectratio]{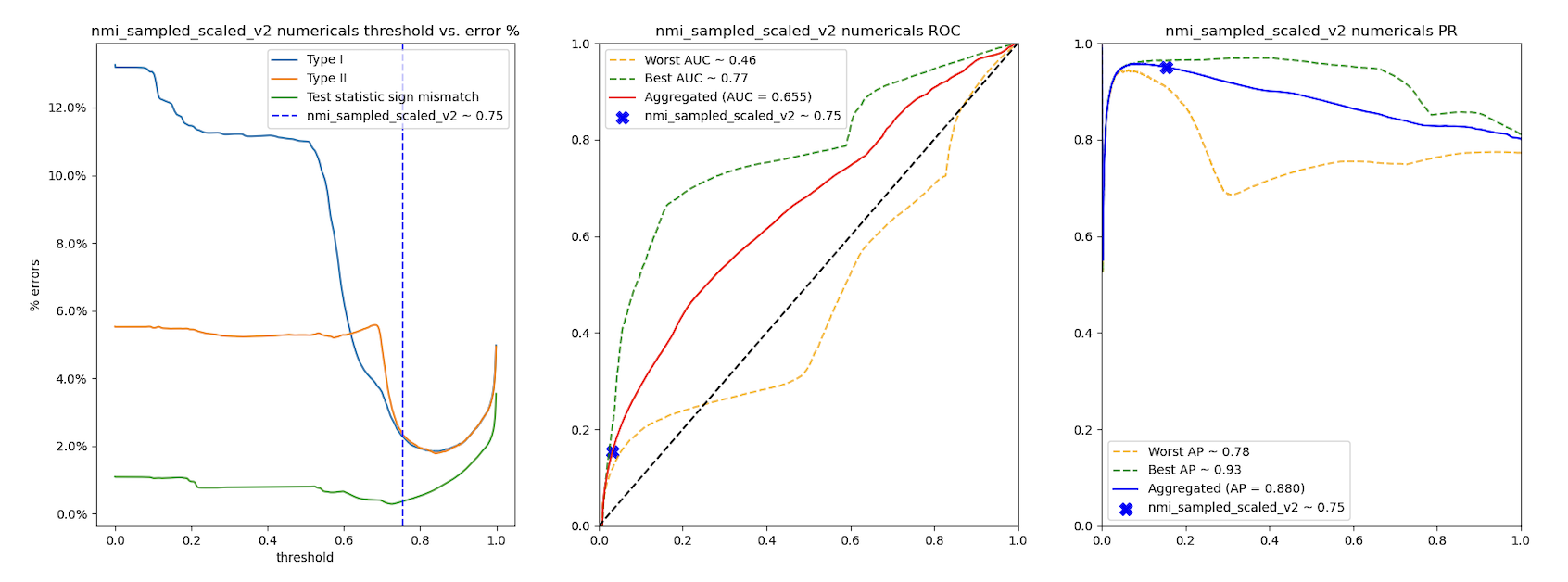}
\caption{Normalized Mutual Information v2 comparison: ROC and Precision-Recall curves for NMIv2 (normalized by anonymized entropy) showing marginally lower performance (AUC 0.66) than NMIv1 (AUC 0.69). Note the non-monotonic error rate behavior with increasing thresholds, invalidating threshold recommendations for NMIv2.}
\label{fig:example2-nmiv2}
\end{figure}

{In this case, NMIv1 has a marginally better efficacy examining the
numbers, e.g. AUC 0.69 \textgreater{} 0.66. In addition we can examine
how the error rate (Total Type I, Type II, test statistic sign mismatch,
divided by total number of statistical tests) varies with respect to the
data quality metric threshold. ~Notice how for NMIv1 the total error
rate (adding the three lines) is non-increasing as the threshold
increases; but for NMIv2 an increase in threshold may yield an
substantive increase in total error rates (so that the algorithm applied
to }{\textit{recommend}}{~an appropriate threshold is invalid and should be
ignored for NMIv2). As discussed in a section above, NMIv2 predominantly
penalizes the injection of entropy, whereas NMIv1 predominantly
penalizes the suppression of entropy. If the anonymization process, when
it injects entropy, does so in a way that doesn't bias the results of
statistical tests, then we don't really care if it injects entropy; but
we do care if it suppresses entropy.}

{}

{Below we provide the overall error rates, for the simulations described
in a prior section, for anonymized datasets that do and do not pass all
the data quality metric minimum thresholds under Project Lighthouse,
i.e. whose minimum data quality metric values do and do not meet the
thresholds above, described in this section:}

{}

{}

\begin{longtable}[]{@{}lllll@{}}
\toprule\noalign{}
\endhead
\bottomrule\noalign{}
\endlastfoot
{} & {Type I} & {Type II} & {Test statistic sign mismatch} & {No
Errors} \\
{Passed DQ thresholds} & {2.78\%} & {1.54\%} & {0.00\%} & {95.68\%} \\
{Did not pass DQ thresholds} & {15.29\%} & {9.85\%} & {0.57\%} &
{74.29\%} \\
\end{longtable}

{}

{}

{}

\section{\texorpdfstring{{Quantitative data minimization for Project
Lighthouse}}{Quantitative data minimization for Project Lighthouse}}\label{h.7iogy51y39wa}

{Despite the extensive security and privacy protections, analysing
users' experiences under Project Lighthouse utilizes Private Data and
thus we consider doing so carefully, and do so following the principle
of data minimization---ie. we utilize as little Private Data as
reasonable to achieve our goal of measuring potential experience gaps,
and no more. This section outlines how we rigorously measure and satisfy
the goal of data minimization---which we term }{quantitative data
minimization}{.}

{}

{For a given analysis, the analyst determines the minimum sample size
required to achieve the analysis goals---this is based on prior analyses
and standard heuristics and analytical methods to determining minimum
detectable effects, ranging from standard analysis best-practice
heuristics to simulation-based power analyses, as in \cite{Basu2020}.
Let this minimum sample size be }{n}{. }

{}

{n}{~is determined with respect to our prior knowledge of the
demographics of Airbnb users in general, and with respect to the
specific statistical tests being used. But it is not determined with
respect to anonymization---in other words, if the privacy controls
outlined in \cite{Basu2020} did not exist, }{n}{~would be the
appropriate sample size to both achieve the goals of the analysis and to
satisfy the principle of data minimization. But we are also imposing the
privacy controls in \cite{Basu2020}, i.e. anonymization. We wish to know
the minimal sample size }{\textgreater= n}{~such that the goals of the
analysis are met---this is precisely the goal, framed as classification
problem, discussed in the empirical justification section above. So that
if all analyses under Project Lighthouse utilize a sample size
\textgreater= }{n}{~whose data quality metrics under anonymization meet
our minimum thresholds we will produce appropriate results in most
cases. We wish to find the minimum such sample size so that we also
achieve the goal of data minimization.}

{}

{In practice we have found that that minimum is usually \textless=
}{2n}{, so that when preparing for an analysis, the analyst produces a
dataset whose size is at-least }{2n}{. With that dataset we run a
sensitivity analysis to find the minimum sample size that is expected to
meet the minimum data quality metric thresholds:}

{}

\begin{enumerate}
\tightlist
\item
  {Select sub-sample sizes in }{{[}n, 2n{]}}{; we do so in roughly
  }{5\%}{~increments.}
\item
  {For each sub-sample size, select }{m}{~sub-samples of that size.}
\item
  {For each sub-sample:}
\end{enumerate}

\begin{enumerate}
\tightlist
\item
  {Anonymize.}
\item
  {Measure data quality using data quality metrics outlined above.}
\item
  {Determine if the minimum data quality metric thresholds are all met.}
\end{enumerate}

{}

{The minimum sample size for the analysis is then the smallest
sub-sample size such that all }{m}{~sub-samples of that size meet all
minimum data quality metric thresholds. As an extra precaution, we fix a
specific sub-sample for analysis and confirm it too meets the minimum
data quality metric thresholds.}

{}

{Because the analyst is an expert in analyzing product experiences at
Airbnb, but not necessarily an expert in anonymization nor private data
analysis, we wish to make this process as easy as possible. We do so
through a guided experience where the analyst runs the sensitivity
analysis, plots the data quality metrics across all sub-samples
analyzed, and determines the minimum---all without the analyst needing
to understand the particulars of anonymization nor data quality metrics.
For instance, below is a plot from a real analysis under Project
Lighthouse, where the minimum sample size to meet data quality metric
thresholds for RILM is 200,000:}

{}

\begin{figure}[H]
\centering
\includegraphics[width=0.8\textwidth,keepaspectratio]{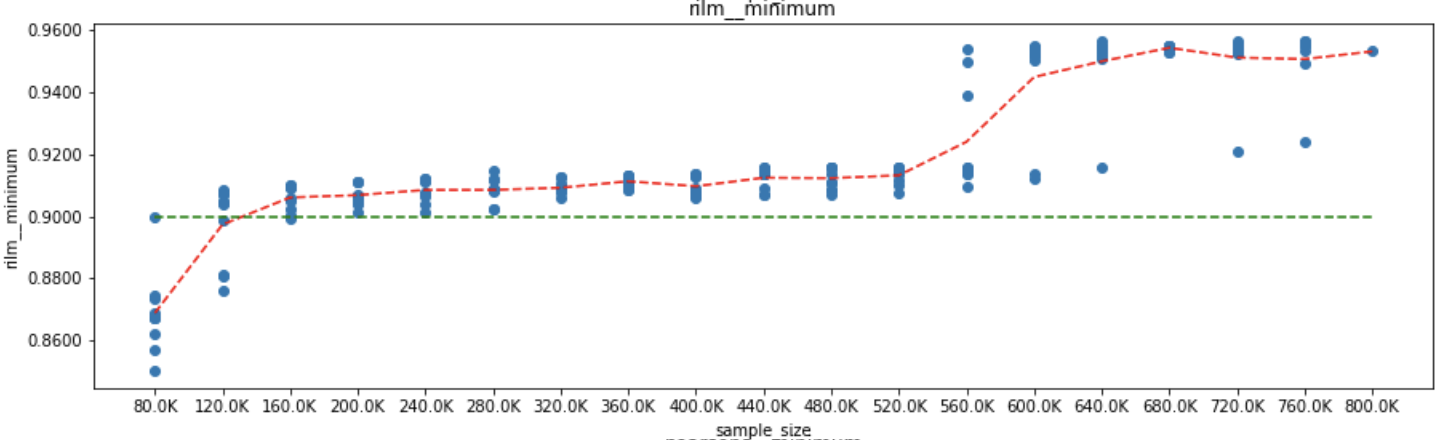}
\caption{Quantitative data minimization sensitivity analysis: Real-world example from Project Lighthouse showing RILM data quality metric values across varying sample sizes. The minimum sample size meeting all data quality thresholds is 200,000, balancing analysis goals with data minimization principles.}
\label{fig:quantitative-data-minimization}
\end{figure}

{Our approach ensures we achieve our goal of measuring potential
experience gaps while following the principle of data minimization.}

\section{\texorpdfstring{{Conclusion}}{Conclusion}}\label{h.dzm8ol99n60m}

{In this followup to \cite{Basu2020}, we describe the data quality
metrics we use to assess data quality under anonymization for Project
Lighthouse. We propose a methodology for assessing data quality metrics
and thresholds, by reframing as a classic machine learning
classification problem. Finally, we show how these metrics and
thresholds are used to rigorously achieve the principle of data
minimization for Project Lighthouse.}

{}

{We have relied on the methods we describe in this paper for many years,
covering a variety of analyses under Project Lighthouse. In sharing
these methods, we hope to help other companies and institutions more
reliably measure data quality under anonymization. In future technical
papers, we will continue to do the same.}

\bibliographystyle{plain}
\bibliography{references}

\end{document}